\documentclass[prd,aps,showpacs,groupedaddress,eqsecnum,notitlepage,nofootinbib,twocolumn,superscriptaddress]{revtex4-2}
\usepackage{graphicx}
\usepackage{tikz,siunitx,mwe}
\usepackage{tensor}
\usepackage{epstopdf}
\usepackage{amsmath}
\usepackage{amsfonts}
\usepackage{slashed}
\usepackage{amssymb}
\usepackage{enumerate, color}
\usepackage{subfigure}
\usepackage{lipsum}
\usepackage{color}
\usepackage{graphicx,bm}
\usepackage[utf8]{inputenc}
\usepackage{eurosym}
\usepackage{scalerel}
\usepackage{float}
\usepackage{accents}
\usepackage[colorlinks=true,pdfstartview=FitV,linkcolor=blue,citecolor=blue,urlcolor=blue,breaklinks=true]{hyperref}
\usepackage{orcidlink}
\newcommand{\be}{\begin{equation}}
\newcommand{\ee}{\end{equation}}
\newcommand{\ben}{\begin{eqnarray}}
\newcommand{\een}{\end{eqnarray}}
\newcommand{\bes}{\begin{subequations}}
\newcommand{\ees}{\end{subequations}}
\usepackage{comment}
\def\bal#1\eal{\begin{align}#1\end{align}}

\newcommand{\sech}{{\rm sech}}
\newcommand{\LL}{{\cal L}}
\newcommand{\Sc}{{\cal S}}
\newcommand{\Hc}{{\cal H}}

\newcommand{\LX}{{\cal L}_X}
\newcommand{\LXX}{{\cal L}_{XX}}
\newcommand{\LXp}{{\cal L}_{X\phi}}

\newcommand{\LP}{{\cal L}_{\phi}}

\newcommand{\nn}{\nonumber\\}
\newcommand{\Nc}{{\cal N}}

\newcommand{\vphi}{{\varphi}}

\newcommand{\n}{\nabla}
\def\bal#1\eal{\begin{align}#1\end{align}}
\begin{document}
\title{Spatially localized scalar structures on hyperscaling violating geometries}
\author{I. Andrade\,\orcidlink{0000-0002-9790-684X}}
        \email[]{andradesigor0@gmail.com}\affiliation{Departamento de F\'\i sica, Universidade Federal da Para\'\i ba, 58051-970 Jo\~ao Pessoa, PB, Brazil}
\author{M.A. Marques\,\orcidlink{0000-0001-7022-5502}}
        \email[]{marques@cbiotec.ufpb.br}\affiliation{Departamento de Biotecnologia, Universidade Federal da Para\'\i ba, 58051-900 Jo\~ao Pessoa, PB, Brazil}
\author{R. Menezes\,\orcidlink{0000-0002-9586-4308}}
     \email[]{rmenezes@dcx.ufpb.br}\affiliation{Departamento de Ci\^encias Exatas, Universidade Federal
da Para\'{\i}ba, 58297-000 Rio Tinto, PB, Brazil}
\author{D.C. Moreira\,\orcidlink{0000-0002-8799-3206}}\email{moreira.dancesar@gmail.com}\affiliation{Centro de Ciências, Tecnologia e Saúde, Universidade Estadual da Paraíba, 58233-000, Araruna, PB, Brazil}
\begin{abstract}
In this work, we investigate probe scalar field models preserving covariance on fixed, static background geometries that present hyperscaling violation properties. We develop a first-order framework that rises from restrictions on the dynamical and hyperscaling violating exponents. The results show that stable, analytical kink-like solutions and their respective energy densities can be obtained for a general class of models. In the canonical model, in particular, these solutions minimize the energy of the system.   
\end{abstract}
\maketitle
\section{Introduction}

Scalar field models are important in several branches of Physics. For instance, they can be used in the study of defect structures in flat spacetime \cite{manton,vachaspati}, such as kinks, vortices and magnetic monopoles, and in curved spacetime, in the braneworld scenarios, where they may be taken in the modeling of the extra dimension \cite{rs2,dewolfe}, or in the search for hairy black hole solutions, where the scalar hair plays a relevant role; see Ref.~\cite{herdeiro2015asymptotically} and references therein. In the context of topological defects, in particular, kinks are spatially localized soliton-like solutions that arise under the action of a single real scalar field in $(1,1)$ spacetime dimensions. The canonical model which supports these structures consists of the difference between the dynamical term (quadratic in the derivatives of the field) and a potential. They, however, may also arise in non-canonical (generalized) models. The first motivation comes from cosmology \cite{a1,a2,a3}, where the generalized models were used to provide a tentative solution for the expansion of the universe at a late stage of its evolution, but over the years several works dealing with non-canonical models that support kink-like solutions appeared in the literature; see, for instance, Refs.~\cite{babichev2006global,sen2003dirac,sen2005tachyon,bazeia2007generalized,bazeia2008first}.

The discussion about searching and obtaining soliton-like scalar field solutions can be extended to setups beyond flat spacetime. In these scenarios, one deals with effective models where probe scalars live on fixed, static background geometries. The problem of the existence and stability of these solutions is addressed as extensions \cite{palmer1979derrick,radmore1978non,carloni2019derrick,mandal2021solitons} and evasions \cite{perivolaropoulos2018gravitational,alestas2019evading,morris2021radially,morris2022bps,moreira2022analytical,moreira2022scalar,moreira2023localized} of Derrick's theorem \cite{derrick1964comments,hobart1963instability} and, as a whole, they suggest that models where the Lagrangian is constituted by probe scalars with standard dynamics can only present stable spatially localized solutions in flat $(1,1)$ dimensional spacetimes and cannot exist in static, asymptotically flat spacetimes in general. Within this context, it is natural to search for analytical well-behaved spatially localized scalar field solutions from fully covariant Lagrangians on non-asymptotically flat backgrounds. The $D$-dimensional Anti-de Sitter spacetime $(AdS_{D})$ is a typical example of such geometries, but other metric setups have emerged in applications of the Anti-de Sitter/Conformal Field Theory $(AdS/CFT)$ correspondence \cite{maldacena1999large,aharony2000large} and its generalization known as gauge/gravity duality \cite{ammon2015gauge}, where is possible to study a series of strongly coupled systems based on analyzes of their gravitational duals. 

Indeed, the studies carried out from the perspective of gauge/gravity duality originally involved calculations on $AdS$ spacetimes due to its conformal invariance, but the need to also deal with non-relativistic models in strongly coupled regimes resulted in the emergence of new geometric backgrounds presenting anisotropic scaling symmetries, where time and space scales differently. The first geometry of this type emerges to describe gravitational duals of field theories that present Lifshitz-type anisotropic scaling \cite{kachru2008gravity} and can be derived from the coupling of a massive vector field to Einstein gravity \cite{taylor2016lifshitz}. It constitutes a class of backgrounds known as Lifshitz spacetimes, characterized by a single parameter known as {\it dynamical exponent}, which acts as a measure of the background anisotropy and departure from the conformal regime. Moreover, a broader class of background geometries that exhibit anisotropic scaling is found within Einstein-Maxwell-dilaton models where, in addition to the dynamical exponent, another parameter called {\it hyperscaling violating exponent} was introduced in order to enable the holographic description of relativistic Fermi liquids, since the emerging black holes in these models have specific heats that can grow linearly with temperature, depending on a choice of parameters \cite{charmousis2010effective,gouteraux2011generalized,dong2012aspects,alishahiha2012charged,taylor2016lifshitz}. The simplest case of these geometries is given by
\begin{eqnarray}\label{lifmetric}\small 
ds^2\!=\!\left(\frac{r}{\ell}\right)^{\!-\frac{2\theta}{D-2}}\!\left(\!-\left(\frac{r}{\ell}\right)^{\!2z}\!\!\!dt^2\!+\!\left(\frac{\ell}{r}\right)^{\!\!2}\! \!dr^2\!+\!\left(\frac{r}{\ell}\right)^{\!2} \!d\vec{x}^{2}\!\right),~~~~~
\end{eqnarray}
where $z$ and $\theta$ denotes the dynamical and hyperscaling violating exponents, respectively. Also, $(x^0,x^1)\equiv (t,r)$, $d\vec{x}^2=dx^kdx^k$, with $k=2, 3, \ldots, D-1$ and, for simplicity, we consider $\ell=1$. In particular, this background is invariant under the set of scaling transformations
\begin{equation}\label{scal}
\!\! t\rightarrow \beta^z t, ~x^k\rightarrow \beta x^k,~ r\rightarrow r/\beta ~~\text{and} ~~ ds\rightarrow \beta^{\frac{\theta}{D-2}}ds.
\end{equation}
The case $\theta=0$ retrieves Lifshitz spacetimes \cite{kachru2008gravity} and for $\left(z,\theta\right)=(1,0)$ one restores the standard $AdS_{D}$ setup.

Since the backgrounds above mentioned have achieved great relevance, the interest in exploring how fields of different natures behave on them is natural. Recently, the study of probe fields on Lifshitz \cite{moreira2022analytical} and asymptotically Lifshitz spacetimes \cite{moreira2022scalar} revealed the existence of regular and radially stable soliton-like scalar field solutions in effective models violating general covariance and the addition of backreactions showed up the existence of asymptotically Lifshitz black holes \cite{moreira2022scalar,Moreira_2024}. Here, we are interested in studying static configurations of spatially localized scalar field solutions from fully covariant models on curved background geometries. Due to the restrictions arising from Derrick's Theorem, geometries presenting anisotropic scaling symmetries appear as natural setups for searching for such solutions.

The investigation deals with a general class of models and covers a wide range of application possibilities. Under a constraint that relates the dynamical and hyperscaling exponents, we present the conditions which leads to first-order equations that enable us to find stable scalar field solutions. In this direction, we first study the canonical model, which supports a BPS-like bound in the energy \cite{bogomol1976stability,prasad1975exact}. We then investigate some non-canonical models, such as the Born-Infeld-like one \cite{brax2003tachyon} and the twinlike one \cite{andrews2010distinguishing,bazeia2011twinlike,bazeia2011new,bazeia2017twinlike}. Using the first-order framework, we show that the scalar field solutions may be mapped into the ones in $(1,1)$ flat spacetime dimensions.

This work is organized as follows. In Sec.~\ref{framework}, we present the general class of models we are interested in and explore its properties by calculating the equation of motion, the energy-momentum tensor and energy. We investigate the conditions to get a first-order equation compatible with the equation of motion, implying a restriction that relates the dynamical and hyperscaling violating exponents. We also investigate the stability of the solutions around small fluctuations, showing that it is governed by a Sturm-Liouville eigenvalue equation which can be factorized in a product of adjoint operators. This ensures that the kink-like solutions found are radially stable. We continue the work in Sec.~\ref{models}, where we present some specific models. First, we study the canonical (or standard) model, where, by using the energy density, we introduce an auxiliary function that allows us to develop the Bogomol'nyi procedure, showing that the solutions of the first-order equations lead to the minimal energy of the system. We illustrate this case for a specific potential, whose minima is connected by the solution. Next, we study a Born-Infeld-like model, showing that a stable field profile similar to the so-called tachyonic kink may appear depending on the exponent value. Then, we investigate two non-canonical models that support the twinlike character. In particular, in the ALTW model \cite{andrews2010distinguishing}, one can only find the conditions to get the same solutions and energy density of the canonical model, as the stability linear differs. We also, based on Refs.~\cite{bazeia2011new,bazeia2017twinlike}, obtain the necessary condition which leads to the same linear stability. We end our work in Sec.~\ref{end}, where the conclusions and some perspectives for future research are presented.

\section{Framework}\label{framework}

In this work we are interested in studying static configurations of classical scalar fields on background geometries that present anisotropic scaling properties. In particular, we address field models with generalized dynamics on the hyperscaling violating metric \eqref{lifmetric}. We consider systems described by the general action
\begin{equation}\label{action}
\Sc=\int d^D x\sqrt{-g}\,\mathcal{L}\left(\phi,X\right),
\end{equation}
where $\phi(x)$ denotes the scalar field and $X = \frac12\nabla_A\phi\nabla^A\phi$ represents its standard dynamics. The form of the above action is motivated by the fact that it encompasses several models of current interest, such as the canonical, Born-Infeld, cuscuton, twinlike and other models, which have been studied in many papers over the years \cite{babichev2006global,sen2003dirac,sen2005tachyon,bazeia2007generalized,brax2003tachyon,bazeia2008first,andrews2010distinguishing,bazeia2009braneworld,bazeia2011twinlike,bazeia2011new,bazeia2017twinlike,cuscuton1,cuscuton2,andrade2019cuscuton}. Here, the scalar field lives on a static $D$-dimensional fixed background geometry with coordinates $x_A$ - with $A=0,1,\cdots, D$ - and whose metric determinant is $g$. In this setup, the scalar field equation becomes
\begin{equation}\label{feom}   \nabla_A\left(\mathcal{L}_X\nabla^A\phi\right)=\mathcal{L}_\phi,
\end{equation}
and  can be expanded as
\be\label{feomexp}
\mathcal{G}^{AB}\n_A\n_B\phi +2X\LXp = \LL_{\phi},
\ee
where we define $\mathcal{G}^{AB}=\LX g^{AB}+\LXX\n^A\phi\n^B\phi$, for simplicity. We have used $\mathcal{L}_\phi=\partial\LL/\partial\phi$, $\mathcal{L}_X =\partial\LL/\partial X$ and $\mathcal{L}_{X\phi} =\partial^2\LL/\partial X\partial\phi$ and so on. The energy-momentum tensor associated to the action \eqref{action} is
\begin{equation}\label{emt}
T_{AB}=g_{AB}\mathcal{L}-\mathcal{L}_X\nabla_A\phi\nabla_A\phi
\end{equation}
and diffeomorphism invariance ensure its conservation -  i.e, $\nabla_{A}T^{AB}=0$. Since static backgrounds have a timelike Killing vector $\xi=-\partial_t$, we can use it to define a conserved current given by $J^A=T^{AB}\xi_B$. Stokes' theorem then allows us to calculate the conserved charge
\begin{equation}\label{consch}
E(\xi)=-\int_\Sigma d^{D-1} x\sqrt{|h|} \eta_A\xi_B T^{AB},
\end{equation}
which is identified as the energy of the field solution on the background geometry. Here, $\eta_A$ denotes the unit
normal vector of the surface $\Sigma$ defined at fixed $t$ and with induced metric $h_{ij}$ such that $\left|h\right|=\text{det}\left(h_{ij}\right)$.

In order to simplify our approach to the problem and take advantage of the role played by the $r$-coordinate, we assume that the scalar field only presents radial dependence - i.e., $\phi=\phi (r)$ - and satisfies the boundary conditions
\be\label{boundcond}
\lim_{r\to \infty}\phi(r)=\phi_{\infty},\qquad\lim_{r\to \infty}\frac{d\phi}{dr}=0,
\ee
where $\phi_{\infty}$ is determined by the model. In this way, the field equation \eqref{feom} becomes
\begin{eqnarray}\label{eom}
\left(\LL -2X\LX\right)^\prime = \frac{2X\LX}{r}\left(\!z\!+\!D\!-\!2 -\frac{D-1}{D-2}\theta\right),~~~~~~
\end{eqnarray}
where primes denote derivation with respect to the radial coordinate. At this point, it is interesting to note that the right-hand side of the equation \eqref{eom} is zero for
\begin{equation}\label{vinc0}
    \frac{\theta}{D-2}=1+\frac{z-1}{D-1},
\end{equation}
which, combined with the boundary conditions at infinity \eqref{boundcond}, leads us to the equation
\begin{equation}\label{stressless}
    \mathcal{L}-2X\mathcal{L}_X=0.
\end{equation}
The above expression reproduces the zero pressure condition, well known from studies of scalar fields in flat space and a strong indicator of linear stability. Therefore, from now on we only deal with setups where the identity \eqref{vinc0} is valid, inducing a constraint on the exponents $z$ and $\theta$.

The expression \eqref{stressless} also allows us to understand the behavior of the solutions at the origin. Since $\mathcal{L}$ gives the energy density, as shown below, it must vanish at $r=0$ to ensure that it is localized. Therefore, at this point, we have that the term $X\mathcal{L}_X$ is zero. For the cases investigated in this work, we have two possibilities. We may have
\be\label{bori1}
\lim_{r\to 0}\phi(r)=\phi_{0},\qquad\lim_{r\to 0}\frac{d\phi}{dr}=0,
\ee
which appears in the canonical model, for instance, or
\be\label{bori2}
\lim_{r\to 0}\phi(r)=\pm\infty,\qquad\lim_{r\to 0}\frac{d\phi}{dr}=\pm\infty,
\ee
which may be present when $\LX\to0$ for $r\to0$. These conditions ensure that the Lagrangian density is regular at the origin and are important to obtain the solutions of interest. The Eq.~\eqref{stressless} also allows us to write the equation of motion as
\be\label{eomstressless}
\left(\LX +2X\LXX\right)\nabla_A\nabla^A\phi = \LP -2X\LXp.
\ee
This equation will be useful in this work.

The energy density of the scalar field is given by
\begin{equation}\label{rho}
    \rho= T_{AB}\xi^{A}\xi^{B}=-\mathcal{L}
\end{equation}
and its associated energy is calculated from Eq.~\eqref{consch}, which leads us to
\be\label{energy}\begin{split}
E\left(\xi\right)&=-\omega_{D-2}\int^\infty_0 dr\frac{\mathcal{L}}{r^{1+\theta_{c}}}\\
    &=-\omega_{D-2}\int^\infty_0 dr\, r^{1+\theta_{c}}\mathcal{L}_X\phi'^2,    
\end{split}
\ee
where, for simplicity, we define $\theta_{c}=\frac{\theta}{D-2}$ and $\omega_{D-2}$ denotes the Euclidean volume related to the $x^i$-coordinates of $\Sigma$. To get the latter line in the above equation, we have used Eq. \eqref{stressless}. We can rewrite the function inside the integral of the above equation as a total derivative by inserting an auxiliary function $W(\phi)$ as follows
\begin{equation}\label{fow}
   -\mathcal{L}_X\phi'= \pm\frac{W_\phi}{r^{1+\theta_{c}}}.
\end{equation}
In this way, the energy is given in terms of the boundary values of the fields, as
\begin{equation}\label{ew}
    E\left(\xi\right)=\left|\Delta W\right|\omega_{D-2},
\end{equation}
with $\Delta W= W(\phi_{\infty})-W(\phi_{0})$.

For static configurations, the standard dynamics of the scalar field becomes $X=r^{2+2\theta_c}{\phi'}^2/2$ and the first-order equation \eqref{fow} allows one to obtain $X$ as a function of $\phi$, in the form $X=F(\phi)$. It must be combined with Eq.~\eqref{stressless} to ensure compatibility with the equation of motion, leading to
\be\label{constraint}
\mathcal{L}\left(\phi,F(\phi)\right) + \sqrt{2F(\phi)}\,W_\phi=0.
\ee
This constraint dictates how the scalar field appears in the Lagrangian density. 

\subsection{Stability}\label{secstab}

We also investigate the linear stability of the model. To do so, we take small fluctuations $\vphi(r,t)$ around the static solution $\phi(r)$ of Eq.~\eqref{eom}, with the field in the form $\phi(r,t)=\phi(r)+\vphi(r,t)$. By substituting this in Eq.~\eqref{feomexp}, one gets
\be
\n_A\left(\mathcal{G}^{AB}\n_B\vphi\right) = \left(\LL_{\phi\phi} -\n_A\left(\LL_{X\phi}\n^A\phi\right)\right)\vphi,
\ee
where $\mathcal{G}^{AB}$ was defined below Eq.~\eqref{feomexp}. We then suppose that the condition \eqref{vinc0} is satisfied, such that the static field $\phi(r)$ obeys Eq.~\eqref{stressless}. Considering that $\vphi(r,t)=\sum_i\vphi_i(r)\cos(\omega_i t)$, the above equation becomes
\be\label{stab}
-\frac{1}{\sigma(r)}\left(\sigma(r)s(r)^2\vphi_i^\prime\right)^\prime +q(r)\vphi_i = \omega_i^2\vphi_i.
\ee
This is a Sturm-Liouville eigenvalue equation in which the weight $\sigma(r)$ and the functions $p(r)$ and $q(r)$ are given by
\bes\label{spq}
\bal
\sigma(r) &= -r^{\left(2D-3\right)\left(1-\theta_c\right)-2}\LX,\\
s(r) &= r^{\left(D-1\right)\left(\theta_c-1\right)+2}\sqrt{\frac{\LX +2X\LXX}{\LX}},\\
q(r) &= \frac{r^{2\left(D-2\right)\left(\theta_c-1\right)}}{\LX}\left(\LL_{\phi\phi} -r^{1+\theta_{c}}\left(r^{1+\theta_{c}}\LL_{X\phi}\phi^\prime\right)^\prime\right).
\eal
\ees
The modes are supposed to be normalized, obeying $\int^\infty_0 \sigma(r)dr\,\vphi_i \vphi_j = \delta_{ij}$. The zero mode $\vphi_0$ is given by
\be
\vphi_0(r) = \Nc r^{1+\theta_{c}}\phi^\prime,
\ee
where $\Nc$ is a normalization constant. The stability equation can be written in the form $L\,\vphi_i = \omega^2_i\,\vphi_i$, where the Sturm-Liouville operator $L$ has the form
\be
L = -\frac{1}{\sigma(r)}\frac{d}{dr}\sigma(r)s(r)^2\frac{d}{dr} +q(r).
\ee
Sturm-Liouville eigenvalue equations \cite{hounkonnou2004factorization} that arise in scalar field models in $(1,1)$ dimensions were previously investigated in Ref.~\cite{andrade2020stability}. Here, the problem is more intricate, as we are in the curved spacetime \cite{moreira2022analytical,moreira2022scalar,moreira2023localized}. Notwithstanding that, the above operator can be factorized in terms of adjoint operators $\Sc_L$ and $\Sc_L^\dagger$, in the form $L = \Sc_L^\dagger \Sc_L$, where
\bes\label{adjointsl}
\bal
\Sc_L &= s\left(-\frac{d}{dr} +\frac{1}{r^{2\left(1+\theta_{c}\right)}\phi^\prime}\left(r^{2\left(1+\theta_{c}\right)}\phi^\prime\right)^\prime\right),\\ 
\Sc_L^\dagger &= s\left(\frac{d}{dr} +\frac{1}{r^{2\left(1+\theta_{c}\right)}\phi^\prime}\left(r^{2\left(1+\theta_{c}\right)}\phi^\prime\right)^\prime\!\!+\!\frac{(\sigma s)^\prime}{\sigma s}\right).~~~~~~~~
\eal
\ees
If these operators are smooth, this factorization ensures that negative eigenvalues are absent in the stability equation \eqref{stab}. Thus, the model is stable against small fluctuations. The Sturm-Liouville eigenvalue equation \eqref{stab} can be transformed into a Schr\"odinger-like one via
\be\label{sltoschrodinger}
dy = -\frac{dr}{s} \quad\text{and}\quad \psi_i = \sqrt{\sigma s}\,\vphi_i.
\ee
The above expressions makes the stability equation being written as $\Hc \psi_i(y) = \omega_i^2 \psi_i(y)$, with $\Hc=-d^2/dy^2 +U(y)$, where the stability potential has the form
\be\label{uschrodinger}
U(y) = \frac{(\sqrt{\sigma s})_{yy}}{\sqrt{\sigma s}}+q(y).
\ee
The subscript $y$ stands for the derivative with respect to $y$. The zero mode is
\be\label{psi0schrodinger}
\psi_0(y) = \Nc r(y)^{1+\theta_{c}}\sqrt{\frac{\sigma(y)}{s(y)}}\,\phi_y.
\ee
The operator $\Hc$ may also be factorized in terms of adjoint operators, in the form $\Hc=\Sc_{\Hc}^\dagger\Sc_{\Hc}$, with
\be
\Sc_{\Hc} = -\frac{d}{dy} +\frac{{\psi_0}_y}{\psi_0} \quad\text{and}\quad \Sc_{\Hc}^\dagger = \frac{d}{dy} +\frac{{\psi_0}_y}{\psi_0}.
\ee
As for the Sturm-Liouville operator, the factorization of $\Hc$ ensures the stability if the above operators are smooth. Next, we illustrate our procedure with some known models.

\section{Models}\label{models}

The first-order framework described in Sec.~\ref{framework} requires that the scalar field in the Lagrangian density obeys a constraint. We shall investigate four possibilities: the standard, Born-Infeld-like \cite{brax2003tachyon}, ALTW \cite{andrews2010distinguishing} and twinlike \cite{bazeia2011new,bazeia2017twinlike} models. Here, for simplicity, we chose to work with cases where $z\geq1$, which implies $\theta\geq D-2$. The case where $z<1$ can be addressed by the same methods.

\subsection{Canonical model}\label{secstandard}
We first study a simpler situation, in which the Lagrangian density describes the canonical model,
\begin{equation}\label{lstandard}
    \mathcal{L}=-X-V(\phi).
\end{equation}
The function $V(\phi)$ represents the potential, which we suppose to support at least two neighbor minima, $v_0$ and $v_\infty$. The equation of motion \eqref{eomstressless} for static solutions reads
\be
r^{2\left(1+\theta_{c}\right)}\left(\phi^{\prime\prime} +\frac{1+\theta_c}{r}\phi^\prime\right) = V_\phi.
\ee
We are interested in obtaining topological solutions, which connect two minima of the potential. Thus, we suppose that $\phi_0=v_0$ and $\phi_\infty=v_\infty$ in the boundary conditions \eqref{bori1} and \eqref{boundcond}. We define the classical mass $m_0$ associated to the minimum $v_0$, via $m_0^2=V_{\phi\phi}|_{v_0}$. For $v_\infty$, we have $m_\infty^2=V_{\phi\phi}|_{v_\infty}$. Near the origin, for $r\approx0$, we can write the solution in the form $\phi(x)=v_0+\phi_{or}(x)$. By substituting it in the above equation, we obtain
\be
\phi_{or}(r) \propto \exp{\left(-\frac{m_0/\theta_c}{r^{\theta_{c}}}\right)}.
\ee
The asymptotic behavior cannot be calculated in general. It depends on the specific model under investigation, as the potential usually engenders infinite classical mass to ensure that the solution connects minima in three spatial dimensions \cite{bazeia2003new}.

The energy density associated to this model is calculated from \eqref{rho}, which becomes
\be\label{rhostdgeral}
\rho(r) = \frac12r^{2\left(1+\theta_{c}\right)}\phi^{\prime 2} +V(\phi).
\ee

Considering that the condition \eqref{vinc0} is satisfied and the line element is \eqref{lifmetric}, we can take advantage of the first-order framework described in Sec.~\ref{framework}. The constraint \eqref{constraint} requires that the potential is written as
\be\label{vwstandard}
V(\phi) = \frac12W_\phi^2,
\ee
such that the model supports the first-order equation \eqref{fow}, which becomes
\be\label{fostd}
\phi^\prime = \pm\frac{W_\phi}{r^{1+\theta_{c}}}.
\ee
The presence of Eqs.~\eqref{vwstandard} and \eqref{fostd} allows us to show that the energy is minimized in the sense of the Bogomol'nyi procedure \cite{bogomol1976stability}. Indeed, by taking the Lagrangian density in Eq.~\eqref{lstandard}, the energy in Eq.~\eqref{energy} can be written as
\be
\begin{aligned}
E &= \omega_{D-2}\int^\infty_0 \frac{dr}{r^{1+\theta_{c}}}\Bigg[\frac12\left(r^{1+\theta_{c}}\phi^\prime \mp W_\phi\right)^2\\
    & \pm r^{1+\theta_{c}}W_\phi\phi^\prime -\frac12W_\phi^2 + V(\phi)\Bigg].
\end{aligned}
\ee
If Eqs.~\eqref{vwstandard} and \eqref{fostd} are satisfied, the energy is minimized, being calculated as in Eq.~\eqref{ew}. In this first-order formalism, the energy density \eqref{rhostdgeral} can be written as
\be\label{rhostd}
\rho(r) = 2V(\phi(r))=W_\phi^2.
\ee

The study of the stability follows the discussion in Sec.~\ref{secstab}. In this case, the functions \eqref{spq} take the form
\bes\label{funcstabstd}\bal \label{sigmastd}
\sigma(r) &= r^{\left(2D-3\right)\left(1-\theta_c\right)-2}, \\ \label{sstd}
s(r) &= r^{\left(D-1\right)\left(\theta_c-1\right)+2},\\ \label{qstd}
q(r) &= r^{2\left(D-2\right)\left(\theta_c-1\right)}V_{\phi\phi}.
\eal
\ees
Notice that $\sigma(r)$ and $s(r)$ do not depend on the potential chosen. The transformation \eqref{sltoschrodinger} can then be obtained explicitly via
\bes
\bal
r(y) &= \Big(\big(1+(D-1)(\theta_c-1)\big)y\Big)^{\frac{-1}{1+\left(D-1\right)\left(\theta_c-1\right)}},\\
\psi_i(y) &= \frac{\vphi_i(r(y))}{r(y)^\frac{\left(D-2\right)\left(\theta_c-1\right)}{2}}.
\eal
\ees
The Schr\"odinger stability potential obtained with this change of variables depends on the specific potential $V(\phi)$ under consideration.

\begin{figure}
    \centering
    \includegraphics[width=8cm]{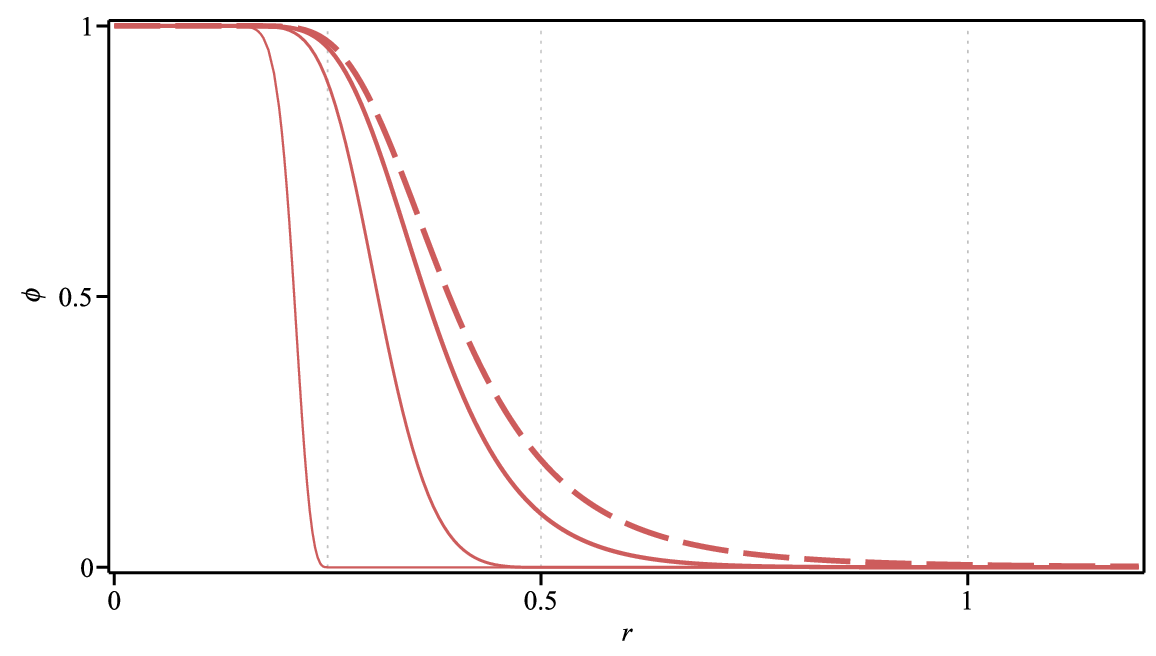}
    \includegraphics[width=8cm]{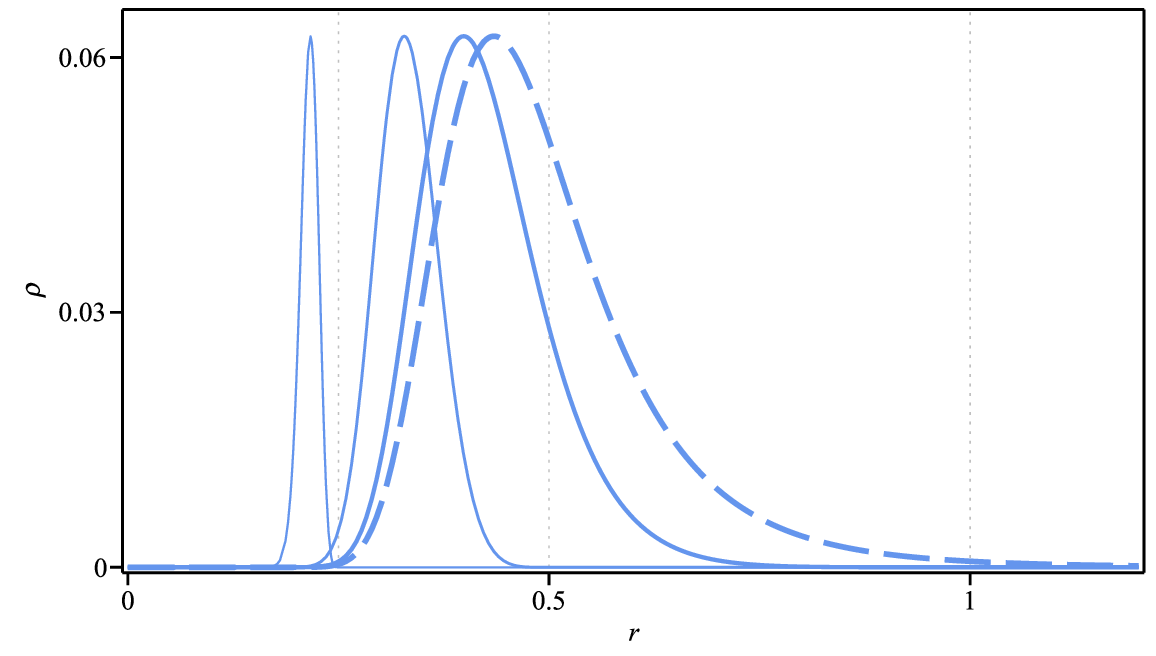}
    \includegraphics[width=8cm]{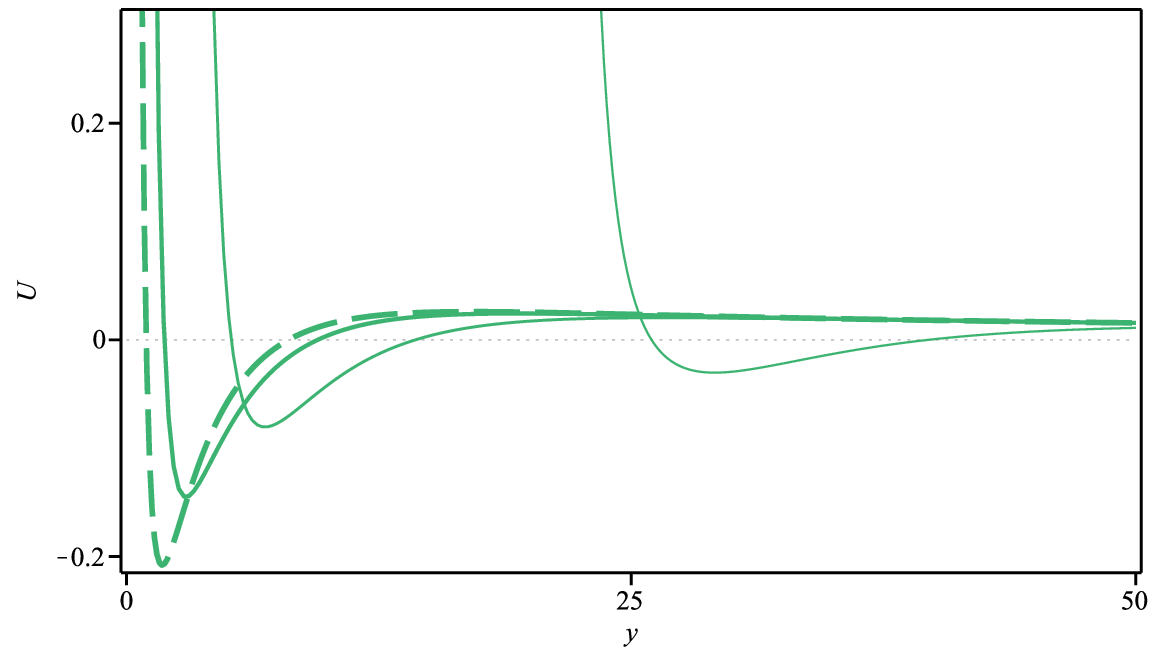}
    \caption{The solution $\phi_+(r)$ in \eqref{phimais} (top), its energy density in Eq.~\eqref{rhostdp} (middle) and the Schr\"odinger-like stability potential $U(y)$ in Eq.~\eqref{ustdp} (bottom) with $D=3$ and $\theta=2$ for $r_0=1/4,1/2, 1$ and the limit $r_0\to+\infty$ (dashed lines). In the stability potential, we have $y=1/(3r^3)$; this makes the potential diverge at the points where $y=1/(3r^3_0)$. The thickness increases with $r_0$. The dotted vertical lines represent the points of compactification, $r=r_0$.}
    \label{figstdp}
\end{figure}

To go further, we need to specify the potential. To ensure that the scalar field solutions connect two neighbor minima of the potential, we take the $p$-model, introduced in Ref.~\cite{bazeia2003new} and also investigated in Refs.~\cite{cruz2009results,cruz2011graviton,mendoncca2015note}, given by
\be
W(\phi) = \frac{p}{2p +1}\phi^{2 +\frac{1}{p}} -\frac{p}{2p -1}\phi^{2 -\frac{1}{p}}.
\ee
The parameter $p$ must be greater than $1$ and odd, i.e., $p=3,5,7,\ldots$, to ensure that the potential engender three degenerate consecutive minima. From Eq.~\eqref{vwstandard}, we see that it has the form
\be
V(\phi) = \frac12\phi^2\left(\phi^\frac{1}{p} -\phi^{-\frac{1}{p}}\right)^2.
\ee
For $p=3,5,7,\ldots$, this potential presents three minima, located at $\phi=0$, with infinite classical mass, and $\phi=\pm1$, with $m_{\pm1}=2/p$. As shown in Ref.~\cite{bazeia2003new}, the presence of a minimum with infinite mass is required to get radially symmetric scalar field solutions connecting two neighbor minima in arbitrary dimensions. Notice that, for $p=1$, the minimum at $\phi=0$ is absent in the potential, so there is no solution connecting two consecutive minima in the geometric background under investigation. Due to this, we exclude the case $p=1$. The first order equation \eqref{fostd} reads
\be
\phi^\prime = \pm \frac{\phi\left(\phi^\frac{1}{p} -\phi^{-\frac{1}{p}}\right)}{r^{1+\theta_{c}}}.
\ee
One may define $x=-\theta_c^{-1}\,r^{-\theta_{c}}$ to get the equation $d\phi/dx = \pm\phi\left(\phi^\frac{1}{p} -\phi^{-\frac{1}{p}}\right)$, which is the same one that appears in the study of kinks in $(1,1)$ flat spacetime dimensions. This trick allows one to find two solutions. The one related to the upper sign in the first-order equation connects the sector $\phi\in [0,1]$ of the potential; it has the form
\be\label{phimais}
\phi_+(r) = \begin{cases}
    \tanh^p\left(\cfrac{1}{p\theta_c}\left(\cfrac{1}{r^{\theta_{c}}} -\cfrac{1}{r_0^{\theta_{c}}}\right)\right), & r\leq r_0\\
    0, & r>r_0,
\end{cases}
\ee
We may also find the solution that connects the sector $\phi\in[-1,0]$ of the potential, which is obtained from the first-order equation with the lower sign; it is given by
\be\label{phimenos}
\phi_-(r) = \begin{cases}
    \tanh^p\left(\cfrac{1}{p\theta_c}\left(\cfrac{1}{r_0^{\theta_{c}}} -\cfrac{1}{r^{\theta_{c}}}\right)\right), & r\leq r_0\\
    0, & r>r_0,
\end{cases}
\ee
where $r_0$ is a constant of integration. Notice that both solutions are compact, with width defined by $\delta = r_0$. The compact solutions connect two neighbor minima of the potential in a compact space. Compact-like solutions have been studied in several papers in the literature \cite{compact1,compact2,compact3,compact4,compact5}. Near the point of compactification, for $r\approx r_0$, we have
\be
{\phi_{\pm}}_{as}(r\approx r0) \approx \pm\left(\frac{1}{p\theta_c}\right)^p\left(\frac{1}{r^{\theta_{c}}} -\frac{1}{r_0^{\theta_{c}}}\right)^p.
\ee
Therefore, as $r_0$ gets larger, the solution gets wider. In the limit $r_0\to\infty$, the solution is not compact anymore, engendering a tail with power-law behavior. In the top panel of Fig.~\ref{figstdp}, we display the solution \eqref{phimais} for some values of $r_0$, including the limit $r_0\to+\infty$ for which the solution is not compact.

The first-order framework allows us to calculate the energy from Eq.~\eqref{ew}, which takes the form
\be
E = \frac{2p}{4p^2 -1}\omega_{D-2},
\ee
for both $\phi_+(x)$ and $\phi_-(x)$ presented above. Notice that the parameter $r_0$ is absent in the above expression. This means that the energy is always the same, regardless its value. Notwithstanding that, the energy density depends on $r_0$, as
\be\label{rhostdp}
\rho(r) = S(r)^4T(r)^{2p-2},
\ee
where
\bes
\bal
S(r) &= \sech\left(\frac{1}{p\theta_c}\left(\frac{1}{r^{\theta_{c}}} -\frac{1}{r_0^{\theta_{c}}}\right)\right),\\
T(r) &= \tanh\left(\frac{1}{p\theta_c}\left(\frac{1}{r^{\theta_{c}}} -\frac{1}{r_0^{\theta_{c}}}\right)\right).
\eal
\ees
This expression is valid for the solutions \eqref{phimais} and \eqref{phimenos}. In the middle panel of Fig.~\ref{figstdp}, we depict the above energy density for some values of $r_0$, including the limit $r_0\to+\infty$ for which the energy density is not compact. 

Moreover, the functions $\sigma(r)$ and $s(r)$ associated to the equations (see Sec.~\ref{secstab}), have the same form in \eqref{sigmastd} and \eqref{sstd}, while $q(r)$ in Eq.~\eqref{qstd} is
\be\begin{aligned}
    q(r) &= \left(1+\frac{1}{p}\right)\left(1+\frac{2}{p}\right)r^{2\left(D-2\right)\left(\theta_c-1\right)}T(r)^2\\
    &+\left(1-\frac{1}{p}\right)\left(1-\frac{2}{p}\right)r^{2\left(D-2\right)\left(\theta_c-1\right)}T(r)^{-2}\\
    &-2r^{2\left(D-2\right)\left(\theta_c-1\right)},
\end{aligned}
\ee
for both solutions. Thus, the adjoint operators \eqref{adjointsl} which factorize the Sturm-Liouville eigenvalue equation are smooth. The zero mode is
\be
\vphi_0(r) = \Nc S(r)^2T(r)^{p-1}.
\ee
As we have explained before, one may also use a Schr\"odinger-like equation to investigate the stability. The stability potential \eqref{uschrodinger} and the zero mode \eqref{psi0schrodinger} are then given by
\bes
\bal\label{ustdp}
U(y)&= \left(1+\frac{1}{p}\right)\left(1+\frac{2}{p}\right)r(y)^{2\left(D-2\right)\left(\theta_c-1\right)}T(r(y))^2\nn
    &+\left(1-\frac{1}{p}\right)\left(1-\frac{2}{p}\right)r(y)^{2\left(D-2\right)\left(\theta_c-1\right)}T(r(y))^{-2}\nn
    &-2r(y)^{2\left(D-2\right)\left(\theta_c-1\right)} -\frac{\vartheta}{y^2}, \\
\psi_0(y)& = \frac{\Nc S(r(y))^2T(r(y))^{p-1}}{r(y)^\frac{\left(D-2\right)\left(\theta_c-1\right)}{2}},
\eal
\ees
where
\be
\vartheta =\frac{\left(D-2\right)^2}{4} \frac{\left(2+D\left(\theta_c-1\right)\right)\left(\theta_c-1\right)}{1+\left(D-1\right)\left(\theta_c-1\right)} .
\ee
In the bottom panel of Fig.~\ref{figstdp}, we display the stability potential \eqref{ustdp} for some values of $r_0$.

\subsection{Born-Infeld-like model}
We continue the investigation considering the Born-Infeld-like model studied in Ref.~\cite{brax2003tachyon}, with
\be\label{lbi}
\LL = -V(\phi)\left(1+2X\right)^a,
\ee
in which $a\geq1/2$ is a parameter that was introduced to generalize the tachyonic dynamics which gives rise to the so-called tachyon kink \cite{sen2003dirac,sen2005tachyon}. As we will show, our procedure allows for the studying of this model on the background geometry \eqref{lifmetric} within a first-order framework. The equation of motion \eqref{eomstressless} associated to the above Lagrangian density for static configurations is
\be
\frac{2ar^{2\left(1+\theta_{c}\right)}(1 +2(2a-1)X)}{(1+2X)(1-2(2a-1)X)}\left(\phi^{\prime\prime} +\frac{1+\theta_c}{r}\phi^\prime\right) = \frac{V_\phi}{V(\phi)}.
\ee
Notice that this equation is much more intricate than the one for the standard model due to the noncanonical nature of the Lagrangian density which describes the model. Thus, the first-order framework allows to simplify the problem.

Let us first analyze the case in which $a>1/2$. From Eq.~\eqref{stressless}, we obtain
\be\label{fobi}
\phi^\prime = \pm\frac{1}{\sqrt{2a-1}r^{1+\theta_{c}}}.
\ee
Interestingly, the latter equation does not depend on the potential $V(\phi)$. It shows that the derivative of the solution diverges at the origin and vanishes asymptotically. The upper/lower sign stands for the increasing/decreasing monotonic solution, which is given by
\be\label{phiapm}
\phi_\pm(r) = \mp\frac{1}{\theta_c\sqrt{2a-1}\,r^{\theta_{c}}} + \overline{\phi},
\ee
where $\overline{\phi}$ is a constant of integration that was chosen to obey $\phi_\pm(+\infty)\to \overline{\phi}$. Also, this solution diverges at the origin. In order to make sense physically, this solution must be analyzed in the context of the energy density \eqref{rho}, which reads
\be\label{rhoaphiprime}
\rho(r) = V(\phi(r))\left(1+r^{2\left(1+\theta_{c}\right)}\,{\phi'}^2\right)^a
\ee
or, by using the first-order equation \eqref{fobi},
\be\label{rhoapm}
       \rho(r) =V(\phi(r))\left(\frac{2a}{2a-1}\right)^a.
\ee
Notice that the behavior of the energy density depends exclusively on $V(\phi(r))$. Therefore, to get the two solutions with finite energy, we impose that $V(\overline{\phi}) = V(\pm\infty)=0$. Using this, the function $W(\phi)$ can be obtained in terms of the potential from Eq.~\eqref{constraint}, in the form
\be\label{vwabi}
V(\phi) = \pm\frac{(2a-1)^{a-\frac12}}{(2a)^a}\frac{dW}{d\phi},
\ee
The upper and lower signs must be taken into account to ensure that the potential is non-negative. To illustrate our procedure, we take a model driven by the function
\be
W(\phi) = \tanh^3(\phi-b+b\tanh(\phi)),
\ee
where $b$ is a non-negative parameter. The potential then becomes
\be\label{potab}
\begin{aligned}
    V(\phi) &= \frac{3(2a-1)^{a-\frac12}}{(2a)^a}\left(1+b\,\sech^2(\phi)\right)\\
    &\hspace{3mm}\sech^2(\phi-b+b\tanh(\phi))\!\tanh^2(\phi-b+b\tanh(\phi)),
\end{aligned}
\ee
which is asymmetric for $b>0$. It exhibits two runaway minima at $\phi\to\pm\infty$ and a minimum at the solution of 
\be
\tilde{\phi}+b\tanh\big(\tilde{\phi}\big)=b.
\ee
To obtain $\tilde{\phi}$, one has to solve this algebraic-transcendental equation numerically, as it does not support analytical solutions for general $b$. To ensure that the solution \eqref{phiapm} connects two neighbor minima of the potential, as required by the boundary conditions \eqref{bori2} and \eqref{boundcond}, we impose that $\overline{\phi}=\tilde{\phi}$. Notice that, for the symmetric potential, when $b=0$, we have $\overline{\phi}=0$.

\begin{figure}
    \centering
    \includegraphics[width=8cm]{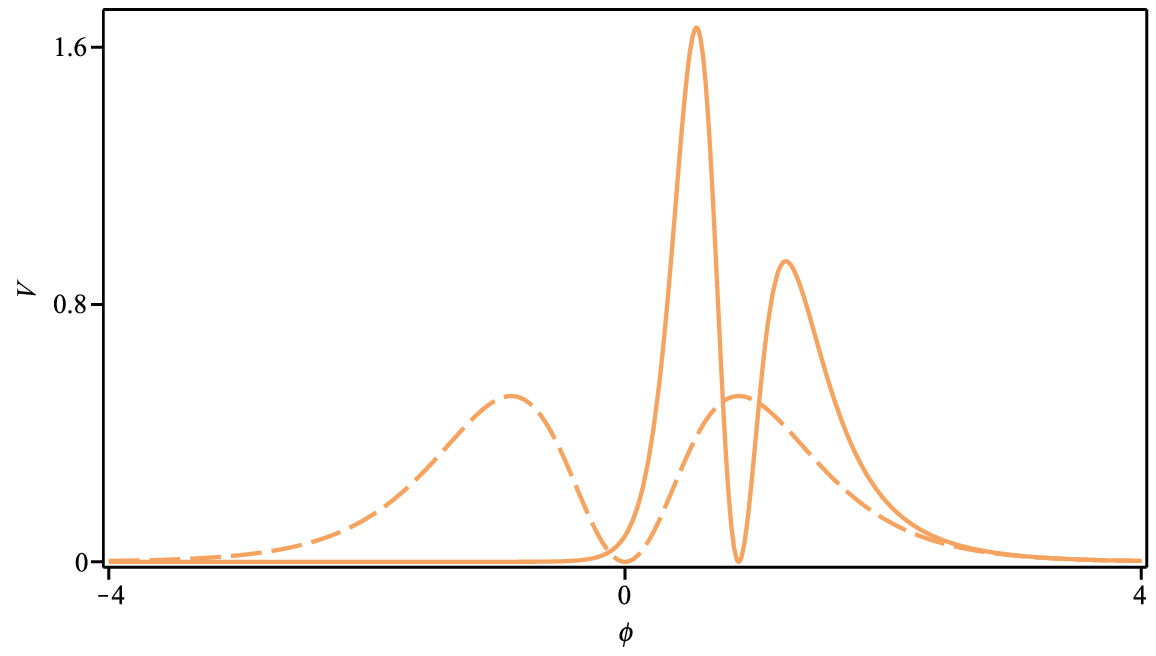}
    \includegraphics[width=8cm]{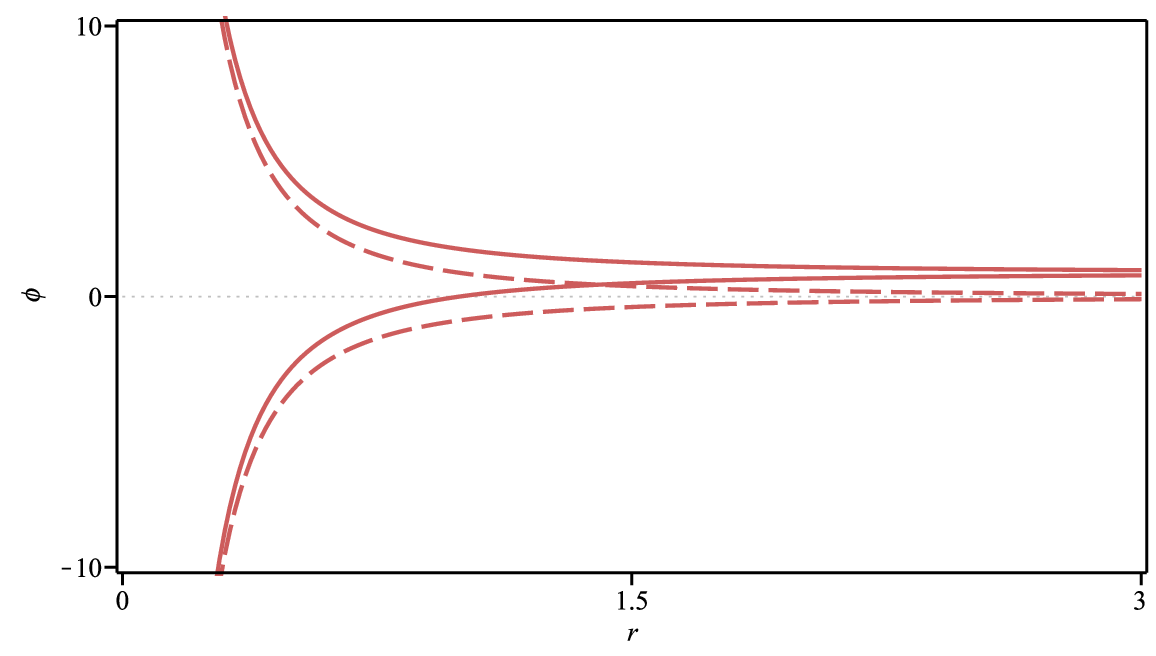}
    \includegraphics[width=8cm]{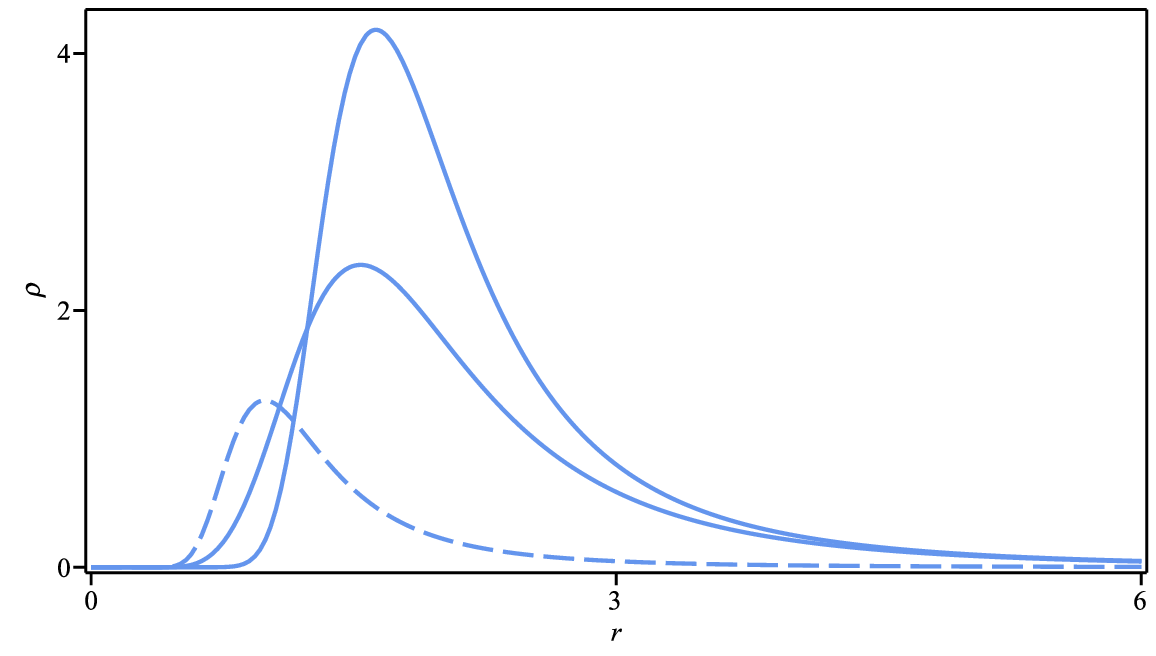}
    \includegraphics[width=8cm]{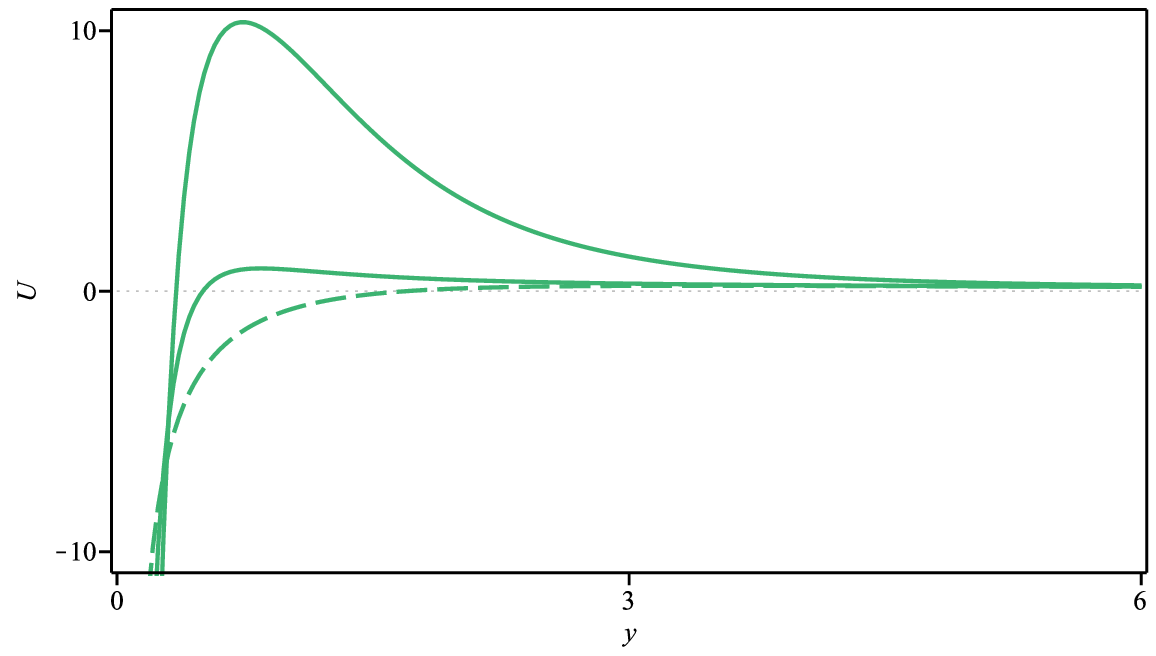}
    \caption{The potential \eqref{potab} (top), the solution $\phi(r)$ in Eq. \eqref{phiapm} (middle-top) and its energy density in Eq.~\eqref{rhoapm} (middle-bottom) and the stability potential \eqref{uabnew} (bottom) with $D=3$, $\theta=2$, $a=2/3$. The dashed lines represent the symmetric case, with $b = 0$ ($\overline{\phi}=0$), and the solid ones stand for $b=3$ $(\overline{\phi} = 0.8803)$. Here, $\phi_+$ is the increasing and $\phi_-$ is the decreasing solution. In the energy density, the line with the highest peak is associated to $\phi_+$, the one right below it is corresponding to $\phi_-$ and the dashed line is for the symmetric solutions. In the stability potential, we have $y=2/(3r^3)$, with the lines following the same logic of the energy density.}
    \label{figbi}
\end{figure}

The potential \eqref{potab} can be combined with the solution \eqref{phiapm} in Eq.~\eqref{rhoapm} to lead to an expression that describes the energy density. It is cumbersome, so we omit it here. We remark that, even though the solution only depends on the parameter $a$, the energy density depends on both $a$ and $b$, since the latter parameter appears in the potential. The energy associated to the solution \eqref{rhoapm} can be calculated from Eq.~\eqref{ew}, which leads us to $E_+=E_- = \omega_{D-2}$. This shows that, despite the solution, potential and energy density are all asymmetric, the energy associated to $\phi_+(r)$ is equal to the one for $\phi_-(r)$. In Fig.~\ref{figbi}, we display the potential \eqref{potab}, the solution $\phi_+$ in Eq.~\eqref{phiapm} and its energy density in Eq.~\eqref{rhoapm}. For the symmetric solutions ($b=0$), the corresponding energy is the same for $\phi_+$ and $\phi_-$ (see the dashed lines in Fig.~\ref{figbi}). For non-null values of $b$, we have that the energy density associated to $\phi_+$ differs from the one related to $\phi_-$. 

The functions \eqref{spq} that drive the stability (see Sec.~\ref{secstab}) are given by
\bes
\bal
\sigma(r) &= (2a)^a(2a-1)^{1-a}r^{\left(2D-3\right)\left(1-\theta_c\right)-2}V(\phi),\\
s(r) &= \left(2-\frac{1}{a}\right)r^{\left(D-1\right)\left(\theta_c-1\right)+2},\\
q(r)&=0.
\eal
\ees
Notice that the functions $s(r)$ and $q(r)$ do not depend on $V(\phi)$. The zero mode is expressed as
\be
\vphi_0(r) = \frac{\Nc}{\sqrt{2a-1}}.
\ee
The change of variables \eqref{sltoschrodinger} that allows to obtain a Schr\"odinger-like equation can be calculated analytically  
\bes
\bal
& r(y) =  \left(\frac{2a-1}{a}\big(1+(D-1)(\theta_c-1)\big)y\right)^{\frac{-1}{1+\left(D-1\right)\left(\theta_c-1\right)}},\\
& \psi_i(y) = \frac{\vphi_i(r(y))}{r(y)^\frac{\left(D-2\right)\left(\theta_c-1\right)}{2}}\sqrt{\frac{2(2a)^aV(\phi(r(y)))}{(2a-1)^{a-2}}}.
\eal
\ees
The stability potential \eqref{uschrodinger} and the zero mode \eqref{psi0schrodinger} for this case are given by
\be\label{uabnew}
U(y) = \frac{\sqrt{r(y)^{(D-2)(1-\theta_c)}V(\phi(r(y)))}_{yy}}{\sqrt{r(y)^{(D-2)(1-\theta_c)}V(\phi(r(y)))}}
\ee
and
\be
\psi_0(y) = \frac{\Nc}{r(y)^\frac{\left(D-2\right)\left(\theta_c-1\right)}{2}}\sqrt{\frac{2(2a)^aV(\phi(r(y)))}{(2a-1)^{a-1}}}.
\ee
The stability potential $U(y)$ is plotted in the bottom panel of Fig.~\ref{figbi}. The asymptotic behavior of $U(y)$ is of a power-law type in $y$. For the specific values taken in the figure, $D=3$, $\theta=2$, $a=2/3$ and both $b = 0$ and $b=3$, it has the form $U(y)\propto y^{-2/3}$, for large values of $y$.

Up to now, we have only dealt with the case $a>1/2$. Now, at this point, we turn our attention to the special case $a=1/2$, for which the Lagrangian density \eqref{lbi} is $\LL = V(\phi)\sqrt{1+2X}$. In this situation, the equation of motion is
\be
\frac{r^{2\left(1+\theta_{c}\right)}}{2(1+X)}\left(\phi^{\prime\prime} +\frac{1+\theta_c}{r}\phi^\prime\right) = \frac{V_\phi}{V(\phi)}.
\ee
To obtain a first order equation, we use \eqref{stressless}, which lead us to
\be\label{stresslesssen}
\frac{V(\phi)}{\sqrt{1+2X}} = 0.
\ee
This equation requires that the slope of the solution is null in points where $V(\phi)\neq0$, as was previously investigated in Refs.~\cite{sen2003dirac,sen2005tachyon}. The tachyonic kinklike solution has the form
\be\label{solbi12}
\phi_\pm(r) = \pm 
\begin{cases}
+\infty,\,\,\,& r<r_0,\\
0,\,\,\,& r=r_0,\\
-\infty, \,\,\, & r>r_0.
\end{cases}
\ee
The energy density can be calculated from Eq.~\eqref{rhoaphiprime} with $a=1/2$. We then write
\be
\rho(r) = V(\phi(r))\sqrt{1+r^{2+2\theta_c}{\phi'}^2}.
\ee
To obtain the value of the energy, we use the method in Ref.~\cite{sen2003dirac} and write $\phi(r)=f(\lambda r)$, where $f$ is an arbitrary function that satisfies $f(0)=-\infty$, $f(r_0)=0$, $f(+\infty)=+\infty$ and $f(\chi)$ is monotonic. The condition \eqref{stresslesssen} requires $\lambda\to\infty$. This trick allows us to write the energy as
\be\label{energybi12}\begin{split}
    E &= \omega_{D-2}\int^{\infty}_{-\infty} dr\,\lambda V(f(\lambda r))\frac{df}{d(\lambda r)}\\
      &= \omega_{D-2}\int^{\infty}_{-\infty} d\chi\,V(\chi),
\end{split}
\ee
where we have taken $\chi=f(\lambda r)$ to obtain the expression in the last line. By taking the limit $a\to1/2$ in Eq.~\eqref{vwabi}, we get that $V(\phi) = \pm W_\phi$. This allows to write the above energy as a surface term, in the form \eqref{ew}, depending only on the boundary values of the field.

We now must provide the potential to get the explicit value of the energy. We must take into account that $V(\phi)$ must be non-negative and have its minima connected by the solution \eqref{solbi12}. To comply with these requirements, we take
\be
W(\phi) = \tanh(\phi) \implies V(\phi) = \sech^2(\phi).
\ee
This potential has a bell shape; it vanishes in the limit $\phi\to\pm\infty$ and presents a maximum at $\phi=0$, such that $V(0)=1$. In this case, the energy from \eqref{energybi12} is $E=2$. This shows that the tachyonic kinklike solution \eqref{solbi12} engenders finite energy. 

\subsection{Twinlike models}
\subsubsection{ALTW Model}
An interesting noncanonical model, which we call ALTW model, was proposed in Ref.~\cite{andrews2010distinguishing}, whose Lagrangian density is
\be\label{laltw}
\LL = M^2 -M^2\sqrt{\left(1 +\frac{2V(\phi)}{M^2}\right)\left(1 +\frac{2X}{M^2}\right)}.
\ee
$M$ is a parameter such that, for large values, the above equation becomes
\be
\LL = -X -V(\phi) +\frac{(X+V(\phi))^2 -4V(\phi)X}{2M^2} +\mathcal{O}\left(\frac{1}{M^4}\right).
\ee
So, in the limit $M^2\to+\infty$, the standard model investigated in Sec.~\ref{secstandard} is recovered.

The equation of motion \eqref{eomstressless} reads
\be
\frac{r^{2\left(1+\theta_{c}\right)}}{M^2+2X}\left(\phi^{\prime\prime} +\frac{1+\theta_c}{r}\phi^\prime\right) = \frac{V_\phi}{M^2+V(\phi)}.
\ee

Following our framework, we can find a first-order equation if the constraint that arises from \eqref{stressless} is satisfied. We then get
\be
M^2\left(1 -\sqrt{\frac{M^2 +2V(\phi)}{M^2 +2X}}\right) = 0.
\ee
The above equation is satisfied for $X=V(\phi)$. This can be compatibilized with Eq.~\eqref{fow} if we take the potential in the form \eqref{vwstandard}. Explicitly, the first-order equation that governs the scalar field is the same in \eqref{fostd}. This makes the energy density being written exactly as in \eqref{rhostd}. Therefore, the model \eqref{laltw} engenders the very same solutions, energy density and energy of the standard model discussed in Sec.~\ref{secstandard}. Since these models share these features, we call them \emph{twinlike} models.

The fact that the two models are twins, however, does not make them indistinguishable, as they reveal different behavior in the study of stability. For the ALTW model, the functions \eqref{spq} associated to the study of the stability (see Sec.~\ref{secstab}) take the form
\bes
\bal
\sigma(r) &= r^{\left(2D-3\right)\left(1-\theta_c\right)-2},\\
s(r) &= \frac{M^2r^{\left(D-1\right)\left(\theta_c-1\right)+2}}{M^2+2V(\phi)},\\
q(r) &= \frac{M^2r^{2\left(D-2\right)\left(\theta_c-1\right)}}{M^2+V(\phi)}\left(V_{\phi\phi} -\frac{2V_\phi^2}{M^2+V(\phi)}\right).
\eal
\ees
By comparing with \eqref{funcstabstd}, we see that these functions are different for a chosen $V(\phi)$. This means that the ALTW and standard models can be distinguished by their linear stability. It is worth commenting that, this differentiation between them tends to disappear as $M^2$ gets larger, as expected, because the limit $M^2\to+\infty$ recovers the standard model.

To study the stability via a Schr\"odinger-like equation, one must perform the change of variables in Eq.~\eqref{sltoschrodinger}. Contrary to the standard model, we were not able to obtain $r(y)$ analytically. So, this investigation must be done numerically. We shall not do it here.  

\subsubsection{Model with strong condition}
The last model which we investigate is described by the Lagrangian density proposed in Ref.~\cite{bazeia2011new} and further studied in Ref.~\cite{bazeia2017twinlike}, given by
\be\label{ltwin}
\LL = -F(Y)V(\phi),\quad\text{with}\quad Y=\frac{X}{V(\phi)}.
\ee
In this model, $F(Y)$ is an arbitrary non-negative function. The $F(Y)=1+Y$ function recovers the standard model. The equation of motion \eqref{eomstressless} takes the form
\be
\frac{\left(F_Y +2F_{YY}Y\right)r^{2\left(1+\theta_{c}\right)}}{F -F_YY +2F_{YY}Y^2}\left(\phi^{\prime\prime} +\frac{1+\theta_c}{r}\phi^\prime\right) = V_\phi.
\ee
We follow the procedure in the previous section and use Eq.~\eqref{stressless}, which reads 
\be
\left(F(Y) -2YF_Y\right)V(\phi) = 0.
\ee
Supposing that this equality holds for all values of $V(\phi)$, we get the algebraic equation $F(Y) -2YF_Y=0$. If this equation does not support real solutions, one cannot obtain twinlike models. Nevertheless, if at least one real positive solution is possible, which we call $Y=c$, we get that $X/V(\phi)=c$, or
\be\label{fotwin}
\phi'= \pm \frac{\sqrt{2cV(\phi)}}{r^{1+\theta_{c}}}.
\ee
We can combine this equation with Eq.~\eqref{fow} to get
\be\label{vwtwin}
F_Y(c)\sqrt{2cV(\phi)} = W_\phi.
\ee
Also, the energy density \eqref{rho} becomes
\be\label{rhotwin}
\rho(r) = F(c)V(\phi(r)).
\ee
We can see from Eqs.~\eqref{fotwin}--\eqref{rhotwin} that, to get the first-order equation \eqref{fostd}, the potential \eqref{vwstandard} and the energy density \eqref{rhostd}, we have to impose the conditions
\bal \label{condtwin}
&c=1, & &F(1)=2, & & F_Y(1)=1.
\eal
These conditions ensure, similarly to the ALTW model, that the noncanonical model \eqref{ltwin} engenders the same solutions, energy density and energy of the standard model. Notice that they restrict the choices for the model, so one must take them into account to provide $F(Y)$. 

In addition, since the conditions to support the twinlike character are imposed on the function $F$ and its derivative, we may investigate how to make the models undistinguishable in their linear stability. In this case, the functions associated to the stability (see Sec.~\ref{secstab}) are
\bes\label{spqtwin2}
\bal
\sigma(r) &= r^{\left(2D-3\right)\left(1-\theta_c\right)-2}F_Y,\\
s(r) &= r^{\left(D-1\right)\left(\theta_c-1\right)+2}\sqrt{\frac{F_Y +2F_{YY}Y}{F_Y}},\\
q(r) &= \frac{r^{2\left(D-2\right)\left(\theta_c-1\right)}}{F_Y}\Bigg(V_{\phi\phi}\left(F-F_YY\right) +\frac{V_\phi^2}{V(\phi)}F_{YY}Y^2\nn
    &~~~~+r^{1+\theta_{c}}\left(\frac{r^{1+\theta_{c}}V_\phi F_{YY}Y\phi^\prime}{V(\phi)}\right)^\prime\Bigg).
\eal
\ees
Substituting the conditions \eqref{condtwin}, these functions become
\bes
\bal
\sigma(r) &= r^{\left(2D-3\right)\left(1-\theta_c\right)-2},\\
s(r) &= r^{\left(D-1\right)\left(\theta_c-1\right)+2}\sqrt{1 +2F_{YY}},\\
q(r) &= r^{2\left(D-2\right)\left(\theta_c-1\right)}\Bigg(V_{\phi\phi} +\frac{V_\phi^2}{V(\phi)}F_{YY}\nn
    &~~~~+\sqrt{2}r^{1+\theta_{c}}\left(\frac{V_\phi F_{YY}}{\sqrt{V(\phi)}}\right)^\prime\Bigg).
\eal
\ees
Notice that, even though the function $\sigma(r)$ takes the same form of the standard model, both $s(r)$ and $q(r)$ are different (see Eq.~\eqref{funcstabstd}). We then see that the conditions \eqref{condtwin} are not sufficient; we have to impose the so-called \emph{strong} condition,
\be\label{strongtwin}
F_{YY}(1)=0,
\ee
in order to get twinlike models in the linear stability.

It is also worth commenting that the change of variables \eqref{sltoschrodinger} cannot be performed analytically using \eqref{spqtwin2} without considering $Y=$ constant, which is the condition that arises from Eq.~\eqref{stressless}. When we impose only the conditions in Eq.~\eqref{condtwin}, we get
\bes
\bal
&r(y)= \Big(\big(1+(D-1)(\theta_c-1)\big)h(1)y\Big)^{\frac{-1}{1+\left(D-1\right)\left(\theta_c-1\right)}},\\
&\psi_i(y) = \frac{\sqrt{h(1)}\vphi_i(r(y))}{r(y)^\frac{\left(D-2\right)\left(\theta_c-1\right)}{2}},
\eal
\ees
where $h(1)=\sqrt{1+2F_{YY}(1)}$. It is straightforward to see that the strong condition, which leads to $h(1)=1$, is necessary to get the twinlike character in the stability. This can be achieved, for instance, by taking $F(Y)=1+Y+(1-Y)^3$.

One may also follow the lines of Ref.~\cite{bazeia2017twinlike} and obtain the conditions under which the models are undistinguishable beyond the linear stability, including terms of the fluctuations with order higher than two. Furthermore, one may also investigate the existence of twinlike models using the Lagrangian density constructed in Ref.~\cite{bazeia2020new}, which allows for the presence of the strong condition and possesses a parameter whose limit leads to the standard model, similarly to the ALTW model.

\section{Ending comments}\label{end}
In this work, we have addressed the problem of obtaining stable spatially localized solutions in fully covariant scalar field models on hyperscaling violating geometries. The capturing of field solutions involves dealing with second-order nonlinear differential equations, which makes the task of calculating analytical solutions very hard, so we have developed a first-order framework, which arises under the condition in Eq.~\eqref{vinc0}. We highlight that the localized solutions found from the first-order equation \eqref{fow} have finite energy and, in the specific case of canonical models, they are compatible with a BPS-like procedure based on energy minimization. By studying the linear stability of the general model, we have found a Sturm-Liouville eigenvalue equation that can be factorized as a product of adjoint operators, ensuring that the kink-like solutions found are stable. We have also obtained the change of variables that allows for the study of the stability via a Schr\"odinger-like equation, with the zero mode explicitly calculated.

To illustrate how our procedure works on several scenarios, we have considered a set of specific models. In the canonical case, we have taken a potential whose neighbor minima are connected by compact solutions. In the Born-Infeld-like model, the $a$ parameter, which appears in the exponent of the Lagrangian density, controls the solution. We have shown that $a>1/2$ leads to a field which connects a runaway to a finite minimum of the potential, but for $a=1/2$, one has a tachyonic kink-like solution, which engenders finite energy even though it has infinite range. 

We have also addressed generalized models which support fields with the same solutions and energy density, so they are called twinlike models. Here, we have considered the ALTW model proposed in \cite{andrews2010distinguishing} to show that it preserves the twinlike feature, with respect to the canonical model, despite differing in the stability structure. We then have shown that the equivalence of the stability structures can be retrieved from the model introduced in \cite{bazeia2011new,bazeia2017twinlike}, which supports additional conditions to obtain the same stability.

As perspectives, in addition to adding backreactions to the present model, one can also look for other background geometries where the formalism developed here can be implemented and new well-behaved solutions in the probe regime can be found. Furthermore, the extension of this formalism to multifield scenarios, in which several scalar fields are present and may be used in the modeling of domain walls, such as the Bloch wall, or to simulate the presence of geometric constrictions \cite{multi1,multi2,multi3,multi4,multi5,multi6}, can also reveal interesting results.

\begin{acknowledgments}
IA, MAM and RM acknowledge financial support from the Brazilian agencies Conselho Nacional de Desenvolvimento Cient\'ifico e Tecnol\'ogico (CNPq), grants 306151/2022-7 (MAM) and 310994/2021-7 (RM), and Paraiba State Research Foundation (FAPESQ-PB) grants 2783/2023 (IA), 0003/2019 (RM) and 0015/2019 (MAM).
\end{acknowledgments}

\bibliography{biblio} 

\begin{thebibliography}{65}%
\makeatletter
\providecommand \@ifxundefined [1]{%
 \@ifx{#1\undefined}
}%
\providecommand \@ifnum [1]{%
 \ifnum #1\expandafter \@firstoftwo
 \else \expandafter \@secondoftwo
 \fi
}%
\providecommand \@ifx [1]{%
 \ifx #1\expandafter \@firstoftwo
 \else \expandafter \@secondoftwo
 \fi
}%
\providecommand \natexlab [1]{#1}%
\providecommand \enquote  [1]{``#1''}%
\providecommand \bibnamefont  [1]{#1}%
\providecommand \bibfnamefont [1]{#1}%
\providecommand \citenamefont [1]{#1}%
\providecommand \href@noop [0]{\@secondoftwo}%
\providecommand \href [0]{\begingroup \@sanitize@url \@href}%
\providecommand \@href[1]{\@@startlink{#1}\@@href}%
\providecommand \@@href[1]{\endgroup#1\@@endlink}%
\providecommand \@sanitize@url [0]{\catcode `\\12\catcode `\$12\catcode `\&12\catcode `\#12\catcode `\^12\catcode `\_12\catcode `\%12\relax}%
\providecommand \@@startlink[1]{}%
\providecommand \@@endlink[0]{}%
\providecommand \url  [0]{\begingroup\@sanitize@url \@url }%
\providecommand \@url [1]{\endgroup\@href {#1}{\urlprefix }}%
\providecommand \urlprefix  [0]{URL }%
\providecommand \Eprint [0]{\href }%
\providecommand \doibase [0]{https://doi.org/}%
\providecommand \selectlanguage [0]{\@gobble}%
\providecommand \bibinfo  [0]{\@secondoftwo}%
\providecommand \bibfield  [0]{\@secondoftwo}%
\providecommand \translation [1]{[#1]}%
\providecommand \BibitemOpen [0]{}%
\providecommand \bibitemStop [0]{}%
\providecommand \bibitemNoStop [0]{.\EOS\space}%
\providecommand \EOS [0]{\spacefactor3000\relax}%
\providecommand \BibitemShut  [1]{\csname bibitem#1\endcsname}%
\let\auto@bib@innerbib\@empty
\bibitem [{\citenamefont {Manton}\ and\ \citenamefont {Sutcliffe}(2004)}]{manton}%
  \BibitemOpen
  \bibfield  {author} {\bibinfo {author} {\bibfnamefont {N.}~\bibnamefont {Manton}}\ and\ \bibinfo {author} {\bibfnamefont {P.}~\bibnamefont {Sutcliffe}},\ }\href@noop {} {\emph {\bibinfo {title} {Topological solitons}}}\ (\bibinfo  {publisher} {Cambridge University Press},\ \bibinfo {year} {2004})\BibitemShut {NoStop}%
\bibitem [{\citenamefont {Alexander}(2000)}]{vachaspati}%
  \BibitemOpen
  \bibfield  {author} {\bibinfo {author} {\bibfnamefont {S.~H.}\ \bibnamefont {Alexander}},\ }\href@noop {} {\emph {\bibinfo {title} {Topological defects in alternative theories to cosmic inflation and string cosmology}}}\ (\bibinfo  {publisher} {Brown University},\ \bibinfo {year} {2000})\BibitemShut {NoStop}%
\bibitem [{\citenamefont {Randall}\ and\ \citenamefont {Sundrum}(1999)}]{rs2}%
  \BibitemOpen
  \bibfield  {author} {\bibinfo {author} {\bibfnamefont {L.}~\bibnamefont {Randall}}\ and\ \bibinfo {author} {\bibfnamefont {R.}~\bibnamefont {Sundrum}},\ }\bibfield  {title} {\bibinfo {title} {An alternative to compactification},\ }\href@noop {} {\bibfield  {journal} {\bibinfo  {journal} {Physical Review Letters}\ }\textbf {\bibinfo {volume} {83}},\ \bibinfo {pages} {4690} (\bibinfo {year} {1999})}\BibitemShut {NoStop}%
\bibitem [{\citenamefont {DeWolfe}\ \emph {et~al.}(2000)\citenamefont {DeWolfe}, \citenamefont {Freedman}, \citenamefont {Gubser},\ and\ \citenamefont {Karch}}]{dewolfe}%
  \BibitemOpen
  \bibfield  {author} {\bibinfo {author} {\bibfnamefont {O.}~\bibnamefont {DeWolfe}}, \bibinfo {author} {\bibfnamefont {D.}~\bibnamefont {Freedman}}, \bibinfo {author} {\bibfnamefont {S.~S.}\ \bibnamefont {Gubser}},\ and\ \bibinfo {author} {\bibfnamefont {A.}~\bibnamefont {Karch}},\ }\bibfield  {title} {\bibinfo {title} {Modeling the fifth dimension with scalars and gravity},\ }\href@noop {} {\bibfield  {journal} {\bibinfo  {journal} {Physical Review D}\ }\textbf {\bibinfo {volume} {62}},\ \bibinfo {pages} {046008} (\bibinfo {year} {2000})}\BibitemShut {NoStop}%
\bibitem [{\citenamefont {Herdeiro}\ and\ \citenamefont {Radu}(2015)}]{herdeiro2015asymptotically}%
  \BibitemOpen
  \bibfield  {author} {\bibinfo {author} {\bibfnamefont {C.~A.}\ \bibnamefont {Herdeiro}}\ and\ \bibinfo {author} {\bibfnamefont {E.}~\bibnamefont {Radu}},\ }\bibfield  {title} {\bibinfo {title} {Asymptotically flat black holes with scalar hair: a review},\ }\href@noop {} {\bibfield  {journal} {\bibinfo  {journal} {International Journal of Modern Physics D}\ }\textbf {\bibinfo {volume} {24}},\ \bibinfo {pages} {1542014} (\bibinfo {year} {2015})}\BibitemShut {NoStop}%
\bibitem [{\citenamefont {Armendariz-Picon}\ \emph {et~al.}(1999)\citenamefont {Armendariz-Picon}, \citenamefont {Damour},\ and\ \citenamefont {Mukhanov}}]{a1}%
  \BibitemOpen
  \bibfield  {author} {\bibinfo {author} {\bibfnamefont {C.}~\bibnamefont {Armendariz-Picon}}, \bibinfo {author} {\bibfnamefont {T.}~\bibnamefont {Damour}},\ and\ \bibinfo {author} {\bibfnamefont {V.}~\bibnamefont {Mukhanov}},\ }\bibfield  {title} {\bibinfo {title} {k-inflation},\ }\href@noop {} {\bibfield  {journal} {\bibinfo  {journal} {Physics Letters B}\ }\textbf {\bibinfo {volume} {458}},\ \bibinfo {pages} {209} (\bibinfo {year} {1999})}\BibitemShut {NoStop}%
\bibitem [{\citenamefont {Armendariz-Picon}\ \emph {et~al.}(2000)\citenamefont {Armendariz-Picon}, \citenamefont {Mukhanov},\ and\ \citenamefont {Steinhardt}}]{a2}%
  \BibitemOpen
  \bibfield  {author} {\bibinfo {author} {\bibfnamefont {C.}~\bibnamefont {Armendariz-Picon}}, \bibinfo {author} {\bibfnamefont {V.}~\bibnamefont {Mukhanov}},\ and\ \bibinfo {author} {\bibfnamefont {P.~J.}\ \bibnamefont {Steinhardt}},\ }\bibfield  {title} {\bibinfo {title} {Dynamical solution to the problem of a small cosmological constant and late-time cosmic acceleration},\ }\href@noop {} {\bibfield  {journal} {\bibinfo  {journal} {Physical Review Letters}\ }\textbf {\bibinfo {volume} {85}},\ \bibinfo {pages} {4438} (\bibinfo {year} {2000})}\BibitemShut {NoStop}%
\bibitem [{\citenamefont {Armendariz-Picon}\ \emph {et~al.}(2001)\citenamefont {Armendariz-Picon}, \citenamefont {Mukhanov},\ and\ \citenamefont {Steinhardt}}]{a3}%
  \BibitemOpen
  \bibfield  {author} {\bibinfo {author} {\bibfnamefont {C.}~\bibnamefont {Armendariz-Picon}}, \bibinfo {author} {\bibfnamefont {V.}~\bibnamefont {Mukhanov}},\ and\ \bibinfo {author} {\bibfnamefont {P.~J.}\ \bibnamefont {Steinhardt}},\ }\bibfield  {title} {\bibinfo {title} {Essentials of k-essence},\ }\href@noop {} {\bibfield  {journal} {\bibinfo  {journal} {Physical Review D}\ }\textbf {\bibinfo {volume} {63}},\ \bibinfo {pages} {103510} (\bibinfo {year} {2001})}\BibitemShut {NoStop}%
\bibitem [{\citenamefont {Babichev}(2006)}]{babichev2006global}%
  \BibitemOpen
  \bibfield  {author} {\bibinfo {author} {\bibfnamefont {E.}~\bibnamefont {Babichev}},\ }\bibfield  {title} {\bibinfo {title} {Global topological k-defects},\ }\href@noop {} {\bibfield  {journal} {\bibinfo  {journal} {Physical Review D}\ }\textbf {\bibinfo {volume} {74}},\ \bibinfo {pages} {085004} (\bibinfo {year} {2006})}\BibitemShut {NoStop}%
\bibitem [{\citenamefont {Sen}(2003)}]{sen2003dirac}%
  \BibitemOpen
  \bibfield  {author} {\bibinfo {author} {\bibfnamefont {A.}~\bibnamefont {Sen}},\ }\bibfield  {title} {\bibinfo {title} {Dirac-{B}orn-{I}nfeld action on the tachyon kink and vortex},\ }\href@noop {} {\bibfield  {journal} {\bibinfo  {journal} {Physical Review D}\ }\textbf {\bibinfo {volume} {68}},\ \bibinfo {pages} {066008} (\bibinfo {year} {2003})}\BibitemShut {NoStop}%
\bibitem [{\citenamefont {Sen}(2005)}]{sen2005tachyon}%
  \BibitemOpen
  \bibfield  {author} {\bibinfo {author} {\bibfnamefont {A.}~\bibnamefont {Sen}},\ }\bibfield  {title} {\bibinfo {title} {Tachyon dynamics in open string theory},\ }\href@noop {} {\bibfield  {journal} {\bibinfo  {journal} {International Journal of Modern Physics A}\ }\textbf {\bibinfo {volume} {20}},\ \bibinfo {pages} {5513} (\bibinfo {year} {2005})}\BibitemShut {NoStop}%
\bibitem [{\citenamefont {Bazeia}\ \emph {et~al.}(2007)\citenamefont {Bazeia}, \citenamefont {Losano}, \citenamefont {Menezes},\ and\ \citenamefont {Oliveira}}]{bazeia2007generalized}%
  \BibitemOpen
  \bibfield  {author} {\bibinfo {author} {\bibfnamefont {D.}~\bibnamefont {Bazeia}}, \bibinfo {author} {\bibfnamefont {L.}~\bibnamefont {Losano}}, \bibinfo {author} {\bibfnamefont {R.}~\bibnamefont {Menezes}},\ and\ \bibinfo {author} {\bibfnamefont {J.}~\bibnamefont {Oliveira}},\ }\bibfield  {title} {\bibinfo {title} {Generalized global defect solutions},\ }\href@noop {} {\bibfield  {journal} {\bibinfo  {journal} {The European Physical Journal C}\ }\textbf {\bibinfo {volume} {51}},\ \bibinfo {pages} {953} (\bibinfo {year} {2007})}\BibitemShut {NoStop}%
\bibitem [{\citenamefont {Bazeia}\ \emph {et~al.}(2008)\citenamefont {Bazeia}, \citenamefont {Losano},\ and\ \citenamefont {Menezes}}]{bazeia2008first}%
  \BibitemOpen
  \bibfield  {author} {\bibinfo {author} {\bibfnamefont {D.}~\bibnamefont {Bazeia}}, \bibinfo {author} {\bibfnamefont {L.}~\bibnamefont {Losano}},\ and\ \bibinfo {author} {\bibfnamefont {R.}~\bibnamefont {Menezes}},\ }\bibfield  {title} {\bibinfo {title} {First-order framework and generalized global defect solutions},\ }\href@noop {} {\bibfield  {journal} {\bibinfo  {journal} {Physics Letters B}\ }\textbf {\bibinfo {volume} {668}},\ \bibinfo {pages} {246} (\bibinfo {year} {2008})}\BibitemShut {NoStop}%
\bibitem [{\citenamefont {Palmer}(1979)}]{palmer1979derrick}%
  \BibitemOpen
  \bibfield  {author} {\bibinfo {author} {\bibfnamefont {T.}~\bibnamefont {Palmer}},\ }\bibfield  {title} {\bibinfo {title} {Derrick's theorem in curved space},\ }\href@noop {} {\bibfield  {journal} {\bibinfo  {journal} {Journal of Physics A: Mathematical and General}\ }\textbf {\bibinfo {volume} {12}},\ \bibinfo {pages} {L17} (\bibinfo {year} {1979})}\BibitemShut {NoStop}%
\bibitem [{\citenamefont {Radmore}\ and\ \citenamefont {Stephenson}(1978)}]{radmore1978non}%
  \BibitemOpen
  \bibfield  {author} {\bibinfo {author} {\bibfnamefont {P.}~\bibnamefont {Radmore}}\ and\ \bibinfo {author} {\bibfnamefont {G.}~\bibnamefont {Stephenson}},\ }\bibfield  {title} {\bibinfo {title} {Non-linear wave equations in a curved background space},\ }\href@noop {} {\bibfield  {journal} {\bibinfo  {journal} {Journal of Physics A: Mathematical and General}\ }\textbf {\bibinfo {volume} {11}},\ \bibinfo {pages} {L149} (\bibinfo {year} {1978})}\BibitemShut {NoStop}%
\bibitem [{\citenamefont {Carloni}\ and\ \citenamefont {Rosa}(2019)}]{carloni2019derrick}%
  \BibitemOpen
  \bibfield  {author} {\bibinfo {author} {\bibfnamefont {S.}~\bibnamefont {Carloni}}\ and\ \bibinfo {author} {\bibfnamefont {J.~L.}\ \bibnamefont {Rosa}},\ }\bibfield  {title} {\bibinfo {title} {Derrick’s theorem in curved spacetime},\ }\href@noop {} {\bibfield  {journal} {\bibinfo  {journal} {Physical Review D}\ }\textbf {\bibinfo {volume} {100}},\ \bibinfo {pages} {025014} (\bibinfo {year} {2019})}\BibitemShut {NoStop}%
\bibitem [{\citenamefont {Mandal}(2021)}]{mandal2021solitons}%
  \BibitemOpen
  \bibfield  {author} {\bibinfo {author} {\bibfnamefont {S.}~\bibnamefont {Mandal}},\ }\bibfield  {title} {\bibinfo {title} {Solitons in curved spacetime},\ }\href@noop {} {\bibfield  {journal} {\bibinfo  {journal} {Europhysics Letters}\ }\textbf {\bibinfo {volume} {136}},\ \bibinfo {pages} {11001} (\bibinfo {year} {2021})}\BibitemShut {NoStop}%
\bibitem [{\citenamefont {Perivolaropoulos}(2018)}]{perivolaropoulos2018gravitational}%
  \BibitemOpen
  \bibfield  {author} {\bibinfo {author} {\bibfnamefont {L.}~\bibnamefont {Perivolaropoulos}},\ }\bibfield  {title} {\bibinfo {title} {Gravitational interactions of finite thickness global topological defects with black holes},\ }\href@noop {} {\bibfield  {journal} {\bibinfo  {journal} {Physical Review D}\ }\textbf {\bibinfo {volume} {97}},\ \bibinfo {pages} {124035} (\bibinfo {year} {2018})}\BibitemShut {NoStop}%
\bibitem [{\citenamefont {Alestas}\ and\ \citenamefont {Perivolaropoulos}(2019)}]{alestas2019evading}%
  \BibitemOpen
  \bibfield  {author} {\bibinfo {author} {\bibfnamefont {G.}~\bibnamefont {Alestas}}\ and\ \bibinfo {author} {\bibfnamefont {L.}~\bibnamefont {Perivolaropoulos}},\ }\bibfield  {title} {\bibinfo {title} {Evading derrick’s theorem in curved space: Static metastable spherical domain wall},\ }\href@noop {} {\bibfield  {journal} {\bibinfo  {journal} {Physical Review D}\ }\textbf {\bibinfo {volume} {99}},\ \bibinfo {pages} {064026} (\bibinfo {year} {2019})}\BibitemShut {NoStop}%
\bibitem [{\citenamefont {Morris}(2021)}]{morris2021radially}%
  \BibitemOpen
  \bibfield  {author} {\bibinfo {author} {\bibfnamefont {J.~R.}\ \bibnamefont {Morris}},\ }\bibfield  {title} {\bibinfo {title} {Radially symmetric scalar solitons},\ }\href@noop {} {\bibfield  {journal} {\bibinfo  {journal} {Physical Review D}\ }\textbf {\bibinfo {volume} {104}},\ \bibinfo {pages} {016013} (\bibinfo {year} {2021})}\BibitemShut {NoStop}%
\bibitem [{\citenamefont {Morris}(2022)}]{morris2022bps}%
  \BibitemOpen
  \bibfield  {author} {\bibinfo {author} {\bibfnamefont {J.~R.}\ \bibnamefont {Morris}},\ }\bibfield  {title} {\bibinfo {title} {{BPS} equations and solutions for maxwell--scalar theory},\ }\href@noop {} {\bibfield  {journal} {\bibinfo  {journal} {Annals of Physics}\ }\textbf {\bibinfo {volume} {438}},\ \bibinfo {pages} {168782} (\bibinfo {year} {2022})}\BibitemShut {NoStop}%
\bibitem [{\citenamefont {Moreira}(2022)}]{moreira2022analytical}%
  \BibitemOpen
  \bibfield  {author} {\bibinfo {author} {\bibfnamefont {D.~C.}\ \bibnamefont {Moreira}},\ }\bibfield  {title} {\bibinfo {title} {Analytical scalar field solutions on {L}ifshitz spacetimes},\ }\href@noop {} {\bibfield  {journal} {\bibinfo  {journal} {Physical Review D}\ }\textbf {\bibinfo {volume} {105}},\ \bibinfo {pages} {016001} (\bibinfo {year} {2022})}\BibitemShut {NoStop}%
\bibitem [{\citenamefont {Moreira}\ \emph {et~al.}(2022)\citenamefont {Moreira}, \citenamefont {Brito},\ and\ \citenamefont {Mota-Silva}}]{moreira2022scalar}%
  \BibitemOpen
  \bibfield  {author} {\bibinfo {author} {\bibfnamefont {D.~C.}\ \bibnamefont {Moreira}}, \bibinfo {author} {\bibfnamefont {F.~A.}\ \bibnamefont {Brito}},\ and\ \bibinfo {author} {\bibfnamefont {J.}~\bibnamefont {Mota-Silva}},\ }\bibfield  {title} {\bibinfo {title} {Scalar fields and {L}ifshitz black holes from {D}errick’s theorem evasion},\ }\href@noop {} {\bibfield  {journal} {\bibinfo  {journal} {Physical Review D}\ }\textbf {\bibinfo {volume} {106}},\ \bibinfo {pages} {125017} (\bibinfo {year} {2022})}\BibitemShut {NoStop}%
\bibitem [{\citenamefont {Moreira}\ \emph {et~al.}(2023)\citenamefont {Moreira}, \citenamefont {Brito},\ and\ \citenamefont {Bazeia}}]{moreira2023localized}%
  \BibitemOpen
  \bibfield  {author} {\bibinfo {author} {\bibfnamefont {D.~C.}\ \bibnamefont {Moreira}}, \bibinfo {author} {\bibfnamefont {F.~A.}\ \bibnamefont {Brito}},\ and\ \bibinfo {author} {\bibfnamefont {D.}~\bibnamefont {Bazeia}},\ }\bibfield  {title} {\bibinfo {title} {Localized scalar structures around static black holes},\ }\href@noop {} {\bibfield  {journal} {\bibinfo  {journal} {Nuclear Physics B}\ }\textbf {\bibinfo {volume} {987}},\ \bibinfo {pages} {116090} (\bibinfo {year} {2023})}\BibitemShut {NoStop}%
\bibitem [{\citenamefont {Derrick}(1964)}]{derrick1964comments}%
  \BibitemOpen
  \bibfield  {author} {\bibinfo {author} {\bibfnamefont {G.}~\bibnamefont {Derrick}},\ }\bibfield  {title} {\bibinfo {title} {Comments on nonlinear wave equations as models for elementary particles},\ }\href@noop {} {\bibfield  {journal} {\bibinfo  {journal} {Journal of Mathematical Physics}\ }\textbf {\bibinfo {volume} {5}},\ \bibinfo {pages} {1252} (\bibinfo {year} {1964})}\BibitemShut {NoStop}%
\bibitem [{\citenamefont {Hobart}(1963)}]{hobart1963instability}%
  \BibitemOpen
  \bibfield  {author} {\bibinfo {author} {\bibfnamefont {R.}~\bibnamefont {Hobart}},\ }\bibfield  {title} {\bibinfo {title} {On the instability of a class of unitary field models},\ }\href@noop {} {\bibfield  {journal} {\bibinfo  {journal} {Proceedings of the Physical Society (1958-1967)}\ }\textbf {\bibinfo {volume} {82}},\ \bibinfo {pages} {201} (\bibinfo {year} {1963})}\BibitemShut {NoStop}%
\bibitem [{\citenamefont {Maldacena}(1999)}]{maldacena1999large}%
  \BibitemOpen
  \bibfield  {author} {\bibinfo {author} {\bibfnamefont {J.}~\bibnamefont {Maldacena}},\ }\bibfield  {title} {\bibinfo {title} {The large-{N} limit of superconformal field theories and supergravity},\ }\href@noop {} {\bibfield  {journal} {\bibinfo  {journal} {International journal of theoretical physics}\ }\textbf {\bibinfo {volume} {38}},\ \bibinfo {pages} {1113} (\bibinfo {year} {1999})}\BibitemShut {NoStop}%
\bibitem [{\citenamefont {Aharony}\ \emph {et~al.}(2000)\citenamefont {Aharony}, \citenamefont {Gubser}, \citenamefont {Maldacena}, \citenamefont {Ooguri},\ and\ \citenamefont {Oz}}]{aharony2000large}%
  \BibitemOpen
  \bibfield  {author} {\bibinfo {author} {\bibfnamefont {O.}~\bibnamefont {Aharony}}, \bibinfo {author} {\bibfnamefont {S.~S.}\ \bibnamefont {Gubser}}, \bibinfo {author} {\bibfnamefont {J.}~\bibnamefont {Maldacena}}, \bibinfo {author} {\bibfnamefont {H.}~\bibnamefont {Ooguri}},\ and\ \bibinfo {author} {\bibfnamefont {Y.}~\bibnamefont {Oz}},\ }\bibfield  {title} {\bibinfo {title} {Large {N} field theories, string theory and gravity},\ }\href@noop {} {\bibfield  {journal} {\bibinfo  {journal} {Physics Reports}\ }\textbf {\bibinfo {volume} {323}},\ \bibinfo {pages} {183} (\bibinfo {year} {2000})}\BibitemShut {NoStop}%
\bibitem [{\citenamefont {Ammon}\ and\ \citenamefont {Erdmenger}(2015)}]{ammon2015gauge}%
  \BibitemOpen
  \bibfield  {author} {\bibinfo {author} {\bibfnamefont {M.}~\bibnamefont {Ammon}}\ and\ \bibinfo {author} {\bibfnamefont {J.}~\bibnamefont {Erdmenger}},\ }\href@noop {} {\emph {\bibinfo {title} {Gauge/gravity duality: Foundations and applications}}}\ (\bibinfo  {publisher} {Cambridge University Press},\ \bibinfo {year} {2015})\BibitemShut {NoStop}%
\bibitem [{\citenamefont {Kachru}\ \emph {et~al.}(2008)\citenamefont {Kachru}, \citenamefont {Liu},\ and\ \citenamefont {Mulligan}}]{kachru2008gravity}%
  \BibitemOpen
  \bibfield  {author} {\bibinfo {author} {\bibfnamefont {S.}~\bibnamefont {Kachru}}, \bibinfo {author} {\bibfnamefont {X.}~\bibnamefont {Liu}},\ and\ \bibinfo {author} {\bibfnamefont {M.}~\bibnamefont {Mulligan}},\ }\bibfield  {title} {\bibinfo {title} {Gravity duals of {L}ifshitz-like fixed points},\ }\href@noop {} {\bibfield  {journal} {\bibinfo  {journal} {Physical Review D}\ }\textbf {\bibinfo {volume} {78}},\ \bibinfo {pages} {106005} (\bibinfo {year} {2008})}\BibitemShut {NoStop}%
\bibitem [{\citenamefont {Taylor}(2016)}]{taylor2016lifshitz}%
  \BibitemOpen
  \bibfield  {author} {\bibinfo {author} {\bibfnamefont {M.}~\bibnamefont {Taylor}},\ }\bibfield  {title} {\bibinfo {title} {{L}ifshitz holography},\ }\href@noop {} {\bibfield  {journal} {\bibinfo  {journal} {Classical and Quantum Gravity}\ }\textbf {\bibinfo {volume} {33}},\ \bibinfo {pages} {033001} (\bibinfo {year} {2016})}\BibitemShut {NoStop}%
\bibitem [{\citenamefont {Charmousis}\ \emph {et~al.}(2010)\citenamefont {Charmousis}, \citenamefont {Gouteraux}, \citenamefont {Soo~Kim}, \citenamefont {Kiritsis},\ and\ \citenamefont {Meyer}}]{charmousis2010effective}%
  \BibitemOpen
  \bibfield  {author} {\bibinfo {author} {\bibfnamefont {C.}~\bibnamefont {Charmousis}}, \bibinfo {author} {\bibfnamefont {B.}~\bibnamefont {Gouteraux}}, \bibinfo {author} {\bibfnamefont {B.}~\bibnamefont {Soo~Kim}}, \bibinfo {author} {\bibfnamefont {E.}~\bibnamefont {Kiritsis}},\ and\ \bibinfo {author} {\bibfnamefont {R.}~\bibnamefont {Meyer}},\ }\bibfield  {title} {\bibinfo {title} {Effective holographic theories for low-temperature condensed matter systems},\ }\href@noop {} {\bibfield  {journal} {\bibinfo  {journal} {Journal of High Energy Physics}\ }\textbf {\bibinfo {volume} {2010}},\ \bibinfo {pages} {1} (\bibinfo {year} {2010})}\BibitemShut {NoStop}%
\bibitem [{\citenamefont {Gout{\'e}raux}\ and\ \citenamefont {Kiritsis}(2011)}]{gouteraux2011generalized}%
  \BibitemOpen
  \bibfield  {author} {\bibinfo {author} {\bibfnamefont {B.}~\bibnamefont {Gout{\'e}raux}}\ and\ \bibinfo {author} {\bibfnamefont {E.}~\bibnamefont {Kiritsis}},\ }\bibfield  {title} {\bibinfo {title} {Generalized holographic quantum criticality at finite density},\ }\href@noop {} {\bibfield  {journal} {\bibinfo  {journal} {Journal of High Energy Physics}\ }\textbf {\bibinfo {volume} {2011}},\ \bibinfo {pages} {1} (\bibinfo {year} {2011})}\BibitemShut {NoStop}%
\bibitem [{\citenamefont {Dong}\ \emph {et~al.}(2012)\citenamefont {Dong}, \citenamefont {Harrison}, \citenamefont {Kachru}, \citenamefont {Torroba},\ and\ \citenamefont {Wang}}]{dong2012aspects}%
  \BibitemOpen
  \bibfield  {author} {\bibinfo {author} {\bibfnamefont {X.}~\bibnamefont {Dong}}, \bibinfo {author} {\bibfnamefont {S.}~\bibnamefont {Harrison}}, \bibinfo {author} {\bibfnamefont {S.}~\bibnamefont {Kachru}}, \bibinfo {author} {\bibfnamefont {G.}~\bibnamefont {Torroba}},\ and\ \bibinfo {author} {\bibfnamefont {H.}~\bibnamefont {Wang}},\ }\bibfield  {title} {\bibinfo {title} {Aspects of holography for theories with hyperscaling violation},\ }\href@noop {} {\bibfield  {journal} {\bibinfo  {journal} {Journal of High Energy Physics}\ }\textbf {\bibinfo {volume} {2012}},\ \bibinfo {pages} {1} (\bibinfo {year} {2012})}\BibitemShut {NoStop}%
\bibitem [{\citenamefont {Alishahiha}\ \emph {et~al.}(2012)\citenamefont {Alishahiha}, \citenamefont {Colg{\'a}in},\ and\ \citenamefont {Yavartanoo}}]{alishahiha2012charged}%
  \BibitemOpen
  \bibfield  {author} {\bibinfo {author} {\bibfnamefont {M.}~\bibnamefont {Alishahiha}}, \bibinfo {author} {\bibfnamefont {E.~{\'O}.}\ \bibnamefont {Colg{\'a}in}},\ and\ \bibinfo {author} {\bibfnamefont {H.}~\bibnamefont {Yavartanoo}},\ }\bibfield  {title} {\bibinfo {title} {Charged black branes with hyperscaling violating factor},\ }\href@noop {} {\bibfield  {journal} {\bibinfo  {journal} {Journal of High Energy Physics}\ }\textbf {\bibinfo {volume} {2012}},\ \bibinfo {pages} {1} (\bibinfo {year} {2012})}\BibitemShut {NoStop}%
\bibitem [{\citenamefont {Moreira}\ \emph {et~al.}(2024)\citenamefont {Moreira}, \citenamefont {Lemos},\ and\ \citenamefont {Brito}}]{Moreira_2024}%
  \BibitemOpen
  \bibfield  {author} {\bibinfo {author} {\bibfnamefont {D.~C.}\ \bibnamefont {Moreira}}, \bibinfo {author} {\bibfnamefont {A.~S.}\ \bibnamefont {Lemos}},\ and\ \bibinfo {author} {\bibfnamefont {F.~A.}\ \bibnamefont {Brito}},\ }\bibfield  {title} {\bibinfo {title} {Charged {L}ifshitz black holes from general covariance breaking},\ }\href@noop {} {\bibfield  {journal} {\bibinfo  {journal} {Classical and Quantum Gravity}\ }\textbf {\bibinfo {volume} {41}},\ \bibinfo {pages} {045004} (\bibinfo {year} {2024})}\BibitemShut {NoStop}%
\bibitem [{\citenamefont {Bogomol'Nyi}(1976)}]{bogomol1976stability}%
  \BibitemOpen
  \bibfield  {author} {\bibinfo {author} {\bibfnamefont {E.}~\bibnamefont {Bogomol'Nyi}},\ }\bibfield  {title} {\bibinfo {title} {The stability of classical solutions},\ }\href@noop {} {\bibfield  {journal} {\bibinfo  {journal} {Sov. J. Nucl. Phys}\ }\textbf {\bibinfo {volume} {24}} (\bibinfo {year} {1976})}\BibitemShut {NoStop}%
\bibitem [{\citenamefont {Prasad}\ and\ \citenamefont {Sommerfield}(1975)}]{prasad1975exact}%
  \BibitemOpen
  \bibfield  {author} {\bibinfo {author} {\bibfnamefont {M.}~\bibnamefont {Prasad}}\ and\ \bibinfo {author} {\bibfnamefont {C.~M.}\ \bibnamefont {Sommerfield}},\ }\bibfield  {title} {\bibinfo {title} {Exact classical solution for the 't {H}ooft monopole and the {J}ulia-{Z}ee dyon},\ }\href@noop {} {\bibfield  {journal} {\bibinfo  {journal} {Physical Review Letters}\ }\textbf {\bibinfo {volume} {35}},\ \bibinfo {pages} {760} (\bibinfo {year} {1975})}\BibitemShut {NoStop}%
\bibitem [{\citenamefont {Brax}\ \emph {et~al.}(2003)\citenamefont {Brax}, \citenamefont {Mourad},\ and\ \citenamefont {Steer}}]{brax2003tachyon}%
  \BibitemOpen
  \bibfield  {author} {\bibinfo {author} {\bibfnamefont {P.}~\bibnamefont {Brax}}, \bibinfo {author} {\bibfnamefont {J.}~\bibnamefont {Mourad}},\ and\ \bibinfo {author} {\bibfnamefont {D.}~\bibnamefont {Steer}},\ }\bibfield  {title} {\bibinfo {title} {Tachyon kinks on non-{BPS} {D}-branes},\ }\href@noop {} {\bibfield  {journal} {\bibinfo  {journal} {Physics Letters B}\ }\textbf {\bibinfo {volume} {575}},\ \bibinfo {pages} {115} (\bibinfo {year} {2003})}\BibitemShut {NoStop}%
\bibitem [{\citenamefont {Andrews}\ \emph {et~al.}(2010)\citenamefont {Andrews}, \citenamefont {Lewandowski}, \citenamefont {Trodden},\ and\ \citenamefont {Wesley}}]{andrews2010distinguishing}%
  \BibitemOpen
  \bibfield  {author} {\bibinfo {author} {\bibfnamefont {M.}~\bibnamefont {Andrews}}, \bibinfo {author} {\bibfnamefont {M.}~\bibnamefont {Lewandowski}}, \bibinfo {author} {\bibfnamefont {M.}~\bibnamefont {Trodden}},\ and\ \bibinfo {author} {\bibfnamefont {D.}~\bibnamefont {Wesley}},\ }\bibfield  {title} {\bibinfo {title} {Distinguishing k-defects from their canonical twins},\ }\href@noop {} {\bibfield  {journal} {\bibinfo  {journal} {Physical Review D}\ }\textbf {\bibinfo {volume} {82}},\ \bibinfo {pages} {105006} (\bibinfo {year} {2010})}\BibitemShut {NoStop}%
\bibitem [{\citenamefont {Bazeia}\ \emph {et~al.}(2011)\citenamefont {Bazeia}, \citenamefont {Dantas}, \citenamefont {Gomes}, \citenamefont {Losano},\ and\ \citenamefont {Menezes}}]{bazeia2011twinlike}%
  \BibitemOpen
  \bibfield  {author} {\bibinfo {author} {\bibfnamefont {D.}~\bibnamefont {Bazeia}}, \bibinfo {author} {\bibfnamefont {J.}~\bibnamefont {Dantas}}, \bibinfo {author} {\bibfnamefont {A.}~\bibnamefont {Gomes}}, \bibinfo {author} {\bibfnamefont {L.}~\bibnamefont {Losano}},\ and\ \bibinfo {author} {\bibfnamefont {R.}~\bibnamefont {Menezes}},\ }\bibfield  {title} {\bibinfo {title} {Twinlike models in scalar field theories},\ }\href@noop {} {\bibfield  {journal} {\bibinfo  {journal} {Physical Review D}\ }\textbf {\bibinfo {volume} {84}},\ \bibinfo {pages} {045010} (\bibinfo {year} {2011})}\BibitemShut {NoStop}%
\bibitem [{\citenamefont {Bazeia}\ and\ \citenamefont {Menezes}(2011)}]{bazeia2011new}%
  \BibitemOpen
  \bibfield  {author} {\bibinfo {author} {\bibfnamefont {D.}~\bibnamefont {Bazeia}}\ and\ \bibinfo {author} {\bibfnamefont {R.}~\bibnamefont {Menezes}},\ }\bibfield  {title} {\bibinfo {title} {New results on twinlike models: Different field theories sharing the same extended solutions},\ }\href@noop {} {\bibfield  {journal} {\bibinfo  {journal} {Physical Review D}\ }\textbf {\bibinfo {volume} {84}},\ \bibinfo {pages} {125018} (\bibinfo {year} {2011})}\BibitemShut {NoStop}%
\bibitem [{\citenamefont {Bazeia}\ \emph {et~al.}(2017)\citenamefont {Bazeia}, \citenamefont {Marques},\ and\ \citenamefont {Menezes}}]{bazeia2017twinlike}%
  \BibitemOpen
  \bibfield  {author} {\bibinfo {author} {\bibfnamefont {D.}~\bibnamefont {Bazeia}}, \bibinfo {author} {\bibfnamefont {M.~A.}\ \bibnamefont {Marques}},\ and\ \bibinfo {author} {\bibfnamefont {R.}~\bibnamefont {Menezes}},\ }\bibfield  {title} {\bibinfo {title} {Twinlike models for kinks, vortices, and monopoles},\ }\href@noop {} {\bibfield  {journal} {\bibinfo  {journal} {Physical Review D}\ }\textbf {\bibinfo {volume} {96}},\ \bibinfo {pages} {025010} (\bibinfo {year} {2017})}\BibitemShut {NoStop}%
\bibitem [{\citenamefont {Bazeia}\ \emph {et~al.}(2009)\citenamefont {Bazeia}, \citenamefont {Gomes}, \citenamefont {Losano},\ and\ \citenamefont {Menezes}}]{bazeia2009braneworld}%
  \BibitemOpen
  \bibfield  {author} {\bibinfo {author} {\bibfnamefont {D.}~\bibnamefont {Bazeia}}, \bibinfo {author} {\bibfnamefont {A.}~\bibnamefont {Gomes}}, \bibinfo {author} {\bibfnamefont {L.}~\bibnamefont {Losano}},\ and\ \bibinfo {author} {\bibfnamefont {R.}~\bibnamefont {Menezes}},\ }\bibfield  {title} {\bibinfo {title} {Braneworld models of scalar fields with generalized dynamics},\ }\href@noop {} {\bibfield  {journal} {\bibinfo  {journal} {Physics Letters B}\ }\textbf {\bibinfo {volume} {671}},\ \bibinfo {pages} {402} (\bibinfo {year} {2009})}\BibitemShut {NoStop}%
\bibitem [{\citenamefont {Afshordi}\ \emph {et~al.}(2007{\natexlab{a}})\citenamefont {Afshordi}, \citenamefont {Chung},\ and\ \citenamefont {Geshnizjani}}]{cuscuton1}%
  \BibitemOpen
  \bibfield  {author} {\bibinfo {author} {\bibfnamefont {N.}~\bibnamefont {Afshordi}}, \bibinfo {author} {\bibfnamefont {D.~J.}\ \bibnamefont {Chung}},\ and\ \bibinfo {author} {\bibfnamefont {G.}~\bibnamefont {Geshnizjani}},\ }\bibfield  {title} {\bibinfo {title} {Causal field theory with an infinite speed of sound},\ }\href@noop {} {\bibfield  {journal} {\bibinfo  {journal} {Physical Review D}\ }\textbf {\bibinfo {volume} {75}},\ \bibinfo {pages} {083513} (\bibinfo {year} {2007}{\natexlab{a}})}\BibitemShut {NoStop}%
\bibitem [{\citenamefont {Afshordi}\ \emph {et~al.}(2007{\natexlab{b}})\citenamefont {Afshordi}, \citenamefont {Chung}, \citenamefont {Doran},\ and\ \citenamefont {Geshnizjani}}]{cuscuton2}%
  \BibitemOpen
  \bibfield  {author} {\bibinfo {author} {\bibfnamefont {N.}~\bibnamefont {Afshordi}}, \bibinfo {author} {\bibfnamefont {D.~J.}\ \bibnamefont {Chung}}, \bibinfo {author} {\bibfnamefont {M.}~\bibnamefont {Doran}},\ and\ \bibinfo {author} {\bibfnamefont {G.}~\bibnamefont {Geshnizjani}},\ }\bibfield  {title} {\bibinfo {title} {Cuscuton cosmology: dark energy meets modified gravity},\ }\href@noop {} {\bibfield  {journal} {\bibinfo  {journal} {Physical Review D}\ }\textbf {\bibinfo {volume} {75}},\ \bibinfo {pages} {123509} (\bibinfo {year} {2007}{\natexlab{b}})}\BibitemShut {NoStop}%
\bibitem [{\citenamefont {Andrade}\ \emph {et~al.}(2019)\citenamefont {Andrade}, \citenamefont {Marques},\ and\ \citenamefont {Menezes}}]{andrade2019cuscuton}%
  \BibitemOpen
  \bibfield  {author} {\bibinfo {author} {\bibfnamefont {I.}~\bibnamefont {Andrade}}, \bibinfo {author} {\bibfnamefont {M.~A.}\ \bibnamefont {Marques}},\ and\ \bibinfo {author} {\bibfnamefont {R.}~\bibnamefont {Menezes}},\ }\bibfield  {title} {\bibinfo {title} {Cuscuton kinks and branes},\ }\href@noop {} {\bibfield  {journal} {\bibinfo  {journal} {Nuclear Physics B}\ }\textbf {\bibinfo {volume} {942}},\ \bibinfo {pages} {188} (\bibinfo {year} {2019})}\BibitemShut {NoStop}%
\bibitem [{\citenamefont {Hounkonnou}\ \emph {et~al.}(2004)\citenamefont {Hounkonnou}, \citenamefont {Sodoga},\ and\ \citenamefont {Azatassou}}]{hounkonnou2004factorization}%
  \BibitemOpen
  \bibfield  {author} {\bibinfo {author} {\bibfnamefont {M.}~\bibnamefont {Hounkonnou}}, \bibinfo {author} {\bibfnamefont {K.}~\bibnamefont {Sodoga}},\ and\ \bibinfo {author} {\bibfnamefont {E.}~\bibnamefont {Azatassou}},\ }\bibfield  {title} {\bibinfo {title} {Factorization of {S}turm--{L}iouville operators: solvable potentials and underlying algebraic structure},\ }\href@noop {} {\bibfield  {journal} {\bibinfo  {journal} {Journal of Physics A: Mathematical and General}\ }\textbf {\bibinfo {volume} {38}},\ \bibinfo {pages} {371} (\bibinfo {year} {2004})}\BibitemShut {NoStop}%
\bibitem [{\citenamefont {Andrade}\ \emph {et~al.}(2020)\citenamefont {Andrade}, \citenamefont {Marques},\ and\ \citenamefont {Menezes}}]{andrade2020stability}%
  \BibitemOpen
  \bibfield  {author} {\bibinfo {author} {\bibfnamefont {I.}~\bibnamefont {Andrade}}, \bibinfo {author} {\bibfnamefont {M.~A.}\ \bibnamefont {Marques}},\ and\ \bibinfo {author} {\bibfnamefont {R.}~\bibnamefont {Menezes}},\ }\bibfield  {title} {\bibinfo {title} {Stability of kinklike structures in generalized models},\ }\href@noop {} {\bibfield  {journal} {\bibinfo  {journal} {Nuclear Physics B}\ }\textbf {\bibinfo {volume} {951}},\ \bibinfo {pages} {114883} (\bibinfo {year} {2020})}\BibitemShut {NoStop}%
\bibitem [{\citenamefont {Bazeia}\ \emph {et~al.}(2003)\citenamefont {Bazeia}, \citenamefont {Menezes},\ and\ \citenamefont {Menezes}}]{bazeia2003new}%
  \BibitemOpen
  \bibfield  {author} {\bibinfo {author} {\bibfnamefont {D.}~\bibnamefont {Bazeia}}, \bibinfo {author} {\bibfnamefont {J.}~\bibnamefont {Menezes}},\ and\ \bibinfo {author} {\bibfnamefont {R.}~\bibnamefont {Menezes}},\ }\bibfield  {title} {\bibinfo {title} {New global defect structures},\ }\href@noop {} {\bibfield  {journal} {\bibinfo  {journal} {Physical review letters}\ }\textbf {\bibinfo {volume} {91}},\ \bibinfo {pages} {241601} (\bibinfo {year} {2003})}\BibitemShut {NoStop}%
\bibitem [{\citenamefont {Cruz}\ \emph {et~al.}(2009)\citenamefont {Cruz}, \citenamefont {Tahim},\ and\ \citenamefont {Almeida}}]{cruz2009results}%
  \BibitemOpen
  \bibfield  {author} {\bibinfo {author} {\bibfnamefont {W.~T.}\ \bibnamefont {Cruz}}, \bibinfo {author} {\bibfnamefont {M.}~\bibnamefont {Tahim}},\ and\ \bibinfo {author} {\bibfnamefont {C.}~\bibnamefont {Almeida}},\ }\bibfield  {title} {\bibinfo {title} {Results in {K}alb-{R}amond field localization and resonances on deformed branes},\ }\href@noop {} {\bibfield  {journal} {\bibinfo  {journal} {Europhysics Letters}\ }\textbf {\bibinfo {volume} {88}},\ \bibinfo {pages} {41001} (\bibinfo {year} {2009})}\BibitemShut {NoStop}%
\bibitem [{\citenamefont {Cruz}\ \emph {et~al.}(2011)\citenamefont {Cruz}, \citenamefont {Gomes},\ and\ \citenamefont {Almeida}}]{cruz2011graviton}%
  \BibitemOpen
  \bibfield  {author} {\bibinfo {author} {\bibfnamefont {W.~T.}\ \bibnamefont {Cruz}}, \bibinfo {author} {\bibfnamefont {A.}~\bibnamefont {Gomes}},\ and\ \bibinfo {author} {\bibfnamefont {C.}~\bibnamefont {Almeida}},\ }\bibfield  {title} {\bibinfo {title} {Graviton resonances on deformed branes},\ }\href@noop {} {\bibfield  {journal} {\bibinfo  {journal} {Europhysics Letters}\ }\textbf {\bibinfo {volume} {96}},\ \bibinfo {pages} {31001} (\bibinfo {year} {2011})}\BibitemShut {NoStop}%
\bibitem [{\citenamefont {Mendon{\c{c}}a}\ and\ \citenamefont {de~Oliveira}(2015)}]{mendoncca2015note}%
  \BibitemOpen
  \bibfield  {author} {\bibinfo {author} {\bibfnamefont {T.}~\bibnamefont {Mendon{\c{c}}a}}\ and\ \bibinfo {author} {\bibfnamefont {H.}~\bibnamefont {de~Oliveira}},\ }\bibfield  {title} {\bibinfo {title} {A note about a new class of two-kinks},\ }\href@noop {} {\bibfield  {journal} {\bibinfo  {journal} {Journal of High Energy Physics}\ }\textbf {\bibinfo {volume} {2015}},\ \bibinfo {pages} {1} (\bibinfo {year} {2015})}\BibitemShut {NoStop}%
\bibitem [{\citenamefont {Flores-Hidalgo}\ and\ \citenamefont {Svaiter}(2002)}]{compact1}%
  \BibitemOpen
  \bibfield  {author} {\bibinfo {author} {\bibfnamefont {G.}~\bibnamefont {Flores-Hidalgo}}\ and\ \bibinfo {author} {\bibfnamefont {N.}~\bibnamefont {Svaiter}},\ }\bibfield  {title} {\bibinfo {title} {Constructing bidimensional scalar field theory models from zero mode fluctuations},\ }\href@noop {} {\bibfield  {journal} {\bibinfo  {journal} {Physical Review D}\ }\textbf {\bibinfo {volume} {66}},\ \bibinfo {pages} {025031} (\bibinfo {year} {2002})}\BibitemShut {NoStop}%
\bibitem [{\citenamefont {Arodz}\ \emph {et~al.}(2005)\citenamefont {Arodz}, \citenamefont {Klimas},\ and\ \citenamefont {Tyranowski}}]{compact2}%
  \BibitemOpen
  \bibfield  {author} {\bibinfo {author} {\bibfnamefont {H.}~\bibnamefont {Arodz}}, \bibinfo {author} {\bibfnamefont {P.}~\bibnamefont {Klimas}},\ and\ \bibinfo {author} {\bibfnamefont {T.}~\bibnamefont {Tyranowski}},\ }\bibfield  {title} {\bibinfo {title} {{Field-theoretic models with {V}-shaped potentials}},\ }\href@noop {} {\bibfield  {journal} {\bibinfo  {journal} {Acta Phys. Polon. B}\ }\textbf {\bibinfo {volume} {36}},\ \bibinfo {pages} {3861} (\bibinfo {year} {2005})}\BibitemShut {NoStop}%
\bibitem [{\citenamefont {Adam}\ \emph {et~al.}(2008)\citenamefont {Adam}, \citenamefont {Grandi}, \citenamefont {Klimas}, \citenamefont {Sanchez-Guillen},\ and\ \citenamefont {Wereszczy{\'n}ski}}]{compact3}%
  \BibitemOpen
  \bibfield  {author} {\bibinfo {author} {\bibfnamefont {C.}~\bibnamefont {Adam}}, \bibinfo {author} {\bibfnamefont {N.}~\bibnamefont {Grandi}}, \bibinfo {author} {\bibfnamefont {P.}~\bibnamefont {Klimas}}, \bibinfo {author} {\bibfnamefont {J.}~\bibnamefont {Sanchez-Guillen}},\ and\ \bibinfo {author} {\bibfnamefont {A.}~\bibnamefont {Wereszczy{\'n}ski}},\ }\bibfield  {title} {\bibinfo {title} {Compact self-gravitating solutions of quartic (k) fields in brane cosmology},\ }\href@noop {} {\bibfield  {journal} {\bibinfo  {journal} {Journal of Physics A: Mathematical and Theoretical}\ }\textbf {\bibinfo {volume} {41}},\ \bibinfo {pages} {375401} (\bibinfo {year} {2008})}\BibitemShut {NoStop}%
\bibitem [{\citenamefont {Arod{\'z}}\ \emph {et~al.}(2008)\citenamefont {Arod{\'z}}, \citenamefont {Klimas},\ and\ \citenamefont {Tyranowski}}]{compact4}%
  \BibitemOpen
  \bibfield  {author} {\bibinfo {author} {\bibfnamefont {H.}~\bibnamefont {Arod{\'z}}}, \bibinfo {author} {\bibfnamefont {P.}~\bibnamefont {Klimas}},\ and\ \bibinfo {author} {\bibfnamefont {T.}~\bibnamefont {Tyranowski}},\ }\bibfield  {title} {\bibinfo {title} {Compact oscillons in the signum-gordon model},\ }\href@noop {} {\bibfield  {journal} {\bibinfo  {journal} {Physical Review D}\ }\textbf {\bibinfo {volume} {77}},\ \bibinfo {pages} {047701} (\bibinfo {year} {2008})}\BibitemShut {NoStop}%
\bibitem [{\citenamefont {Bazeia}\ \emph {et~al.}(2014)\citenamefont {Bazeia}, \citenamefont {Losano}, \citenamefont {Marques},\ and\ \citenamefont {Menezes}}]{compact5}%
  \BibitemOpen
  \bibfield  {author} {\bibinfo {author} {\bibfnamefont {D.}~\bibnamefont {Bazeia}}, \bibinfo {author} {\bibfnamefont {L.}~\bibnamefont {Losano}}, \bibinfo {author} {\bibfnamefont {M.}~\bibnamefont {Marques}},\ and\ \bibinfo {author} {\bibfnamefont {R.}~\bibnamefont {Menezes}},\ }\bibfield  {title} {\bibinfo {title} {Compact structures in standard field theory},\ }\href@noop {} {\bibfield  {journal} {\bibinfo  {journal} {Europhysics Letters}\ }\textbf {\bibinfo {volume} {107}},\ \bibinfo {pages} {61001} (\bibinfo {year} {2014})}\BibitemShut {NoStop}%
\bibitem [{\citenamefont {Bazeia}\ \emph {et~al.}(2020{\natexlab{a}})\citenamefont {Bazeia}, \citenamefont {Losano}, \citenamefont {Marques},\ and\ \citenamefont {Menezes}}]{bazeia2020new}%
  \BibitemOpen
  \bibfield  {author} {\bibinfo {author} {\bibfnamefont {D.}~\bibnamefont {Bazeia}}, \bibinfo {author} {\bibfnamefont {L.}~\bibnamefont {Losano}}, \bibinfo {author} {\bibfnamefont {M.~A.}\ \bibnamefont {Marques}},\ and\ \bibinfo {author} {\bibfnamefont {R.}~\bibnamefont {Menezes}},\ }\bibfield  {title} {\bibinfo {title} {New twinlike models for scalar fields},\ }\href@noop {} {\bibfield  {journal} {\bibinfo  {journal} {Europhysics Letters}\ }\textbf {\bibinfo {volume} {131}},\ \bibinfo {pages} {31002} (\bibinfo {year} {2020}{\natexlab{a}})}\BibitemShut {NoStop}%
\bibitem [{\citenamefont {Bazeia}\ \emph {et~al.}(1995)\citenamefont {Bazeia}, \citenamefont {Dos~Santos},\ and\ \citenamefont {Ribeiro}}]{multi1}%
  \BibitemOpen
  \bibfield  {author} {\bibinfo {author} {\bibfnamefont {D.}~\bibnamefont {Bazeia}}, \bibinfo {author} {\bibfnamefont {M.}~\bibnamefont {Dos~Santos}},\ and\ \bibinfo {author} {\bibfnamefont {R.}~\bibnamefont {Ribeiro}},\ }\bibfield  {title} {\bibinfo {title} {Solitons in systems of coupled scalar fields},\ }\href@noop {} {\bibfield  {journal} {\bibinfo  {journal} {Physics Letters A}\ }\textbf {\bibinfo {volume} {208}},\ \bibinfo {pages} {84} (\bibinfo {year} {1995})}\BibitemShut {NoStop}%
\bibitem [{\citenamefont {Bazeia}\ \emph {et~al.}(1997)\citenamefont {Bazeia}, \citenamefont {Nascimento}, \citenamefont {Ribeiro},\ and\ \citenamefont {Toledo}}]{multi2}%
  \BibitemOpen
  \bibfield  {author} {\bibinfo {author} {\bibfnamefont {D.}~\bibnamefont {Bazeia}}, \bibinfo {author} {\bibfnamefont {J.}~\bibnamefont {Nascimento}}, \bibinfo {author} {\bibfnamefont {R.}~\bibnamefont {Ribeiro}},\ and\ \bibinfo {author} {\bibfnamefont {D.}~\bibnamefont {Toledo}},\ }\bibfield  {title} {\bibinfo {title} {Soliton stability in systems of two real scalar fields},\ }\href@noop {} {\bibfield  {journal} {\bibinfo  {journal} {Journal of Physics A: Mathematical and General}\ }\textbf {\bibinfo {volume} {30}},\ \bibinfo {pages} {8157} (\bibinfo {year} {1997})}\BibitemShut {NoStop}%
\bibitem [{\citenamefont {Izquierdo}\ \emph {et~al.}(2002)\citenamefont {Izquierdo}, \citenamefont {Leon},\ and\ \citenamefont {Guilarte}}]{multi3}%
  \BibitemOpen
  \bibfield  {author} {\bibinfo {author} {\bibfnamefont {A.~A.}\ \bibnamefont {Izquierdo}}, \bibinfo {author} {\bibfnamefont {M.~G.}\ \bibnamefont {Leon}},\ and\ \bibinfo {author} {\bibfnamefont {J.~M.}\ \bibnamefont {Guilarte}},\ }\bibfield  {title} {\bibinfo {title} {Kink variety in systems of two coupled scalar fields in two space-time dimensions},\ }\href@noop {} {\bibfield  {journal} {\bibinfo  {journal} {Physical Review D}\ }\textbf {\bibinfo {volume} {65}},\ \bibinfo {pages} {085012} (\bibinfo {year} {2002})}\BibitemShut {NoStop}%
\bibitem [{\citenamefont {Bazeia}\ \emph {et~al.}(2002)\citenamefont {Bazeia}, \citenamefont {Losano},\ and\ \citenamefont {Wotzasek}}]{multi4}%
  \BibitemOpen
  \bibfield  {author} {\bibinfo {author} {\bibfnamefont {D.}~\bibnamefont {Bazeia}}, \bibinfo {author} {\bibfnamefont {L.}~\bibnamefont {Losano}},\ and\ \bibinfo {author} {\bibfnamefont {C.}~\bibnamefont {Wotzasek}},\ }\bibfield  {title} {\bibinfo {title} {Domain walls in three-field models},\ }\href@noop {} {\bibfield  {journal} {\bibinfo  {journal} {Physical Review D}\ }\textbf {\bibinfo {volume} {66}},\ \bibinfo {pages} {105025} (\bibinfo {year} {2002})}\BibitemShut {NoStop}%
\bibitem [{\citenamefont {Bazeia}\ \emph {et~al.}(2020{\natexlab{b}})\citenamefont {Bazeia}, \citenamefont {Liao},\ and\ \citenamefont {Marques}}]{multi5}%
  \BibitemOpen
  \bibfield  {author} {\bibinfo {author} {\bibfnamefont {D.}~\bibnamefont {Bazeia}}, \bibinfo {author} {\bibfnamefont {M.~A.}\ \bibnamefont {Liao}},\ and\ \bibinfo {author} {\bibfnamefont {M.~A.}\ \bibnamefont {Marques}},\ }\bibfield  {title} {\bibinfo {title} {Geometrically constrained kinklike configurations},\ }\href@noop {} {\bibfield  {journal} {\bibinfo  {journal} {The European Physical Journal Plus}\ }\textbf {\bibinfo {volume} {135}},\ \bibinfo {pages} {1} (\bibinfo {year} {2020}{\natexlab{b}})}\BibitemShut {NoStop}%
\bibitem [{\citenamefont {Bazeia}\ \emph {et~al.}(2022)\citenamefont {Bazeia}, \citenamefont {Marques},\ and\ \citenamefont {Paganelly}}]{multi6}%
  \BibitemOpen
  \bibfield  {author} {\bibinfo {author} {\bibfnamefont {D.}~\bibnamefont {Bazeia}}, \bibinfo {author} {\bibfnamefont {M.~A.}\ \bibnamefont {Marques}},\ and\ \bibinfo {author} {\bibfnamefont {M.}~\bibnamefont {Paganelly}},\ }\bibfield  {title} {\bibinfo {title} {Manipulating the internal structure of {B}loch walls},\ }\href@noop {} {\bibfield  {journal} {\bibinfo  {journal} {The European Physical Journal Plus}\ }\textbf {\bibinfo {volume} {137}},\ \bibinfo {pages} {1} (\bibinfo {year} {2022})}\BibitemShut {NoStop}%
\end{thebibliography}%
\end{document}